\documentclass[aip, reprint, floatfix]{revtex4-1}
\usepackage{graphicx}
\usepackage{dcolumn}
\usepackage{bm}
\usepackage{microtype}
\usepackage[utf8]{inputenc}
\usepackage[T1]{fontenc}
\usepackage{mathptmx}
\usepackage{etoolbox}
\usepackage{array}    
\usepackage{booktabs} 
\usepackage[table,xcdraw]{xcolor} 

\definecolor{lightgray}{RGB}{230, 230, 230}  

\makeatletter
\def\@email#1#2{%
 \endgroup
 \patchcmd{\titleblock@produce}
  {\frontmatter@RRAPformat}
  {\frontmatter@RRAPformat{\produce@RRAP{*#1\href{mailto:#2}{#2}}}\frontmatter@RRAPformat}
  {}{}
}%
\makeatother

\renewcommand\thesection{\Roman{section}} 

\renewcommand\thesubsection{\Alph{subsection}}

\renewcommand\thesubsubsection{\thesubsection.\arabic{subsubsection}}

\begin{document}

\preprint{AIP/123-QED}

\title{Multiscale simulation and machine learning facilitated design of two-dimensional nanomaterials-based tunnel field-effect transistors: a review }
\author{Chloe Isabella Tsang}
 \affiliation{Pritzker School of Molecular Engineering, University of Chicago, Chicago, Illinois 60637,United States.
}
\author{Haihui Pu}
\affiliation{Pritzker School of Molecular Engineering, University of Chicago, Chicago, Illinois 60637,United States.
}
\affiliation{Chemical Sciences and Engineering Division, Physical Sciences and Engineering Directorate, Argonne National Laboratory, Lemont, Illinois 60439, United States.}
\author{Junhong Chen*}
 \email{junhongchen@uchicago.edu}
 \affiliation{Pritzker School of Molecular Engineering, University of Chicago, Chicago, Illinois 60637,United States.
 }
 \affiliation{Chemical Sciences and Engineering Division, Physical Sciences and Engineering Directorate, Argonne National Laboratory, Lemont, Illinois 60439, United States.}

\date{\today}

\begin{abstract}
 Traditional transistors based on complementary metal-oxide-semiconductor (CMOS) and metal-oxide-semiconductor field-effect transistors (MOSFETs) are facing significant limitations as device scaling reaches the limits of Moore's Law. These limitations include increased leakage currents, pronounced short-channel effects (SCEs), and quantum tunneling through the gate oxide, leading to higher power consumption and deviations from ideal behavior. Tunnel Field-Effect Transistors (TFETs) can overcome these challenges by utilizing quantum tunneling of charge carriers to switch between on and off states and achieve a subthreshold swing (SS) below 60 mV/decade. This allows for lower power consumption, continued scaling, and improved performance in low-power applications. This review focuses on the design and operation of TFETs, emphasizing the optimization of device performance through material selection and advanced simulation techniques. The discussion will specifically address the use of two-dimensional (2D) materials in TFET design and explore simulation methods ranging from multi-scale (MS) approaches to machine learning (ML)-driven optimization.
\end{abstract}

\maketitle

\section{Introduction}
Despite the impressive performance of traditional CMOS and MOSFET in the modern electronics industry, the acceleration of technological advancement is beginning to challenge the limits of Moore's law.\cite{7878935, 6186749, yang2014memristive} Consequently, alternative transistor devices\cite{5706338,angelov2019technology, kim2024future, 7878935} are being investigated with the objective of overcoming the performance issues that devices such as MOSFETs and CMOS are prone to present.\cite{1388765} The optimal characteristics of a high-performing device include a small SS, low power consumption, minimal leakage current, and a high on/off current ratio. \cite{kotlyar2013bandgap} Nevertheless, the intrinsic switching mechanism of conventional semiconductor devices represents a significant limitation.\cite{vashchenko2008physical, shi2021review} The operation of traditional semiconductor devices is based on p-n carrier transport, which restricts the SS of these devices to a value exceeding 60 mV/dec.\cite{ionescu_tunnel_2011} Moreover, this also constrains their capacity to achieve reduced power consumption. 

By operating on a different switching mechanism, namely band-to-band tunneling (BTBT), TFETs can achieve a sub-thermal threshold swing of less than 60 mV/dec and gain access to a lower switching power.\cite{avci_comparison_2011, 7465689, yirak2023operation, 8805123, 8919391} Consequently, TFETs have been the subject of considerable research and development as a potential solution to the limitations of conventional FET-based devices. However, this same mechanism that enables these favorable characteristics also presents a trade-off between a low SS and a high on-current.\cite{llorente2019new} The presence of indirect band gaps and a low tunneling probability frequently results in the observation of relatively low on-current levels.\cite{8374834} It is therefore crucial to consider the various aspects of TFET design, including material systems, supply and threshold voltages, device geometries, and other factors that can potentially impact the SS. The material system plays a pivotal role in optimizing the tunneling rate and reducing SS. In particular, heterojunctions\cite{dewey_iiix2013v_2012} are of greater interest than homojunctions,\cite{7409811} as single-material-based devices are unable to accommodate steep band profiles due to the presence of different doping levels \cite{ionescu_tunnel_2011}. The use of heterojunctions allows for the integration of disparate materials at the source and channel, thereby facilitating the formation of an abrupt interface.\cite{oldham1964interface} This, in turn, permits a reduction in the width of the tunnel barrier and an increase in the tunneling probability.\cite{seabaugh_low-voltage_2010} The nature of TFET devices requires that the band edges be sharply defined at the interfaces. Accordingly, the design of the source-channel junction has a significant impact on the device's performance.\cite{lionti_area-selective_2022} MS modeling\cite{strangio2017benchmarks} of the TFET device is an indispensable component of the design optimization process. This requires the use of sophisticated softwares (e.g.,Quantum Espresso, Wannier90, and NanoTCAD ViDES.)\cite{fiori_multiscale_2013} The flexibility and adaptability of computational simulation are particularly advantageous in the identification and resolution of single-crystal defects and other design issues, as well as in the mitigation of other atomistic issues.

This review will address the salient features of TFET design and prediction, with a particular emphasis on 2D heterojunction devices and the most prevalent materials utilized in their fabrication. The primary distinctions between the FET and TFET device will be elucidated in Section II. Section III presents a comprehensive analysis of various 2D material systems and notable heterojunctions that have demonstrated significant potential for TFET device design. In the forth section, a variety of TFET material systems are examined, their simulated performances are detailed, and their potential for meeting the International Roadmap for Devices and Systems (IDRS) target requirements for high-performance future digital applications is discussed. The simulation of these devices is based on density functional theory (DFT) calculations, which predict the electronic structure and thermodynamic properties of channel materials. The calculation of a given material system's fundamental properties provides the basis for subsequent model development, enabling the prediction of device transfer characteristics. Section V examines the role of ML-based methods in facilitating the discovery of novel heterojunction materials.

\section{Key Differences Between FETs and TFETs}
The fundamental distinction between MOSFETs and TFETs can be attributed to their disparate carrier transport and switching methodologies.\cite{https://doi.org/10.1002/adma.202106975} To fully comprehend this, it is essential to initially acknowledge that a MOSFET is a barrier-controlled device.\cite{arora2007mosfet} The application of a gate voltage is necessary to raise or lower the potential energy barrier between the source and drain. \cite{colinge2008finfets} Once the energy barrier has been reduced to an adequate level, electrons from the source are able to traverse through the channel and reach the drain via thermionic emission (Fig.1A).\cite{kanungo_2d_2022} This method of carrier transport presents significant challenges for the MOSFET downscaling process, which is otherwise known as SCEs.\cite{taur_fundamentals_2009} A reduction in the channel length of the device results in a decrease in effective doping, which in turn leads to a lowering of the threshold voltage(V\textsubscript{T}). This correlation between threshold voltage and channel length provides insight into one of the primary SCEs,\cite{colinge2008finfets} namely, drain-induced barrier lowering (DIBL).\cite{1485954} The application of a positive voltage to the drain results in a reduction in both the overall channel length and the threshold voltage of the device. This is due to the fact that the applied drain voltage results in an increase in the depletion layer, which in turn reduces the overall channel length and threshold voltage.\cite{taur_fundamentals_2009} Furthermore, this SCE gives rise to a variation in the subthreshold current with high drain biases. A number of studies have examined potential solutions to this phenomenon and have identified several promising avenues for addressing it. These include reducing the thickness of the oxide,\cite{taur_fundamentals_2009} increasing the doping concentration of the substrate, and exploring alternative doping methods,\cite{kumawat2023tunnel} such as halo\cite{arefinia2009novel,kumar2023design, akturk2003increased} and pocket doping\cite{5640662,6942145, bhuyan2020review}.

\begin{figure*}
    \centering\includegraphics[width=1\linewidth]{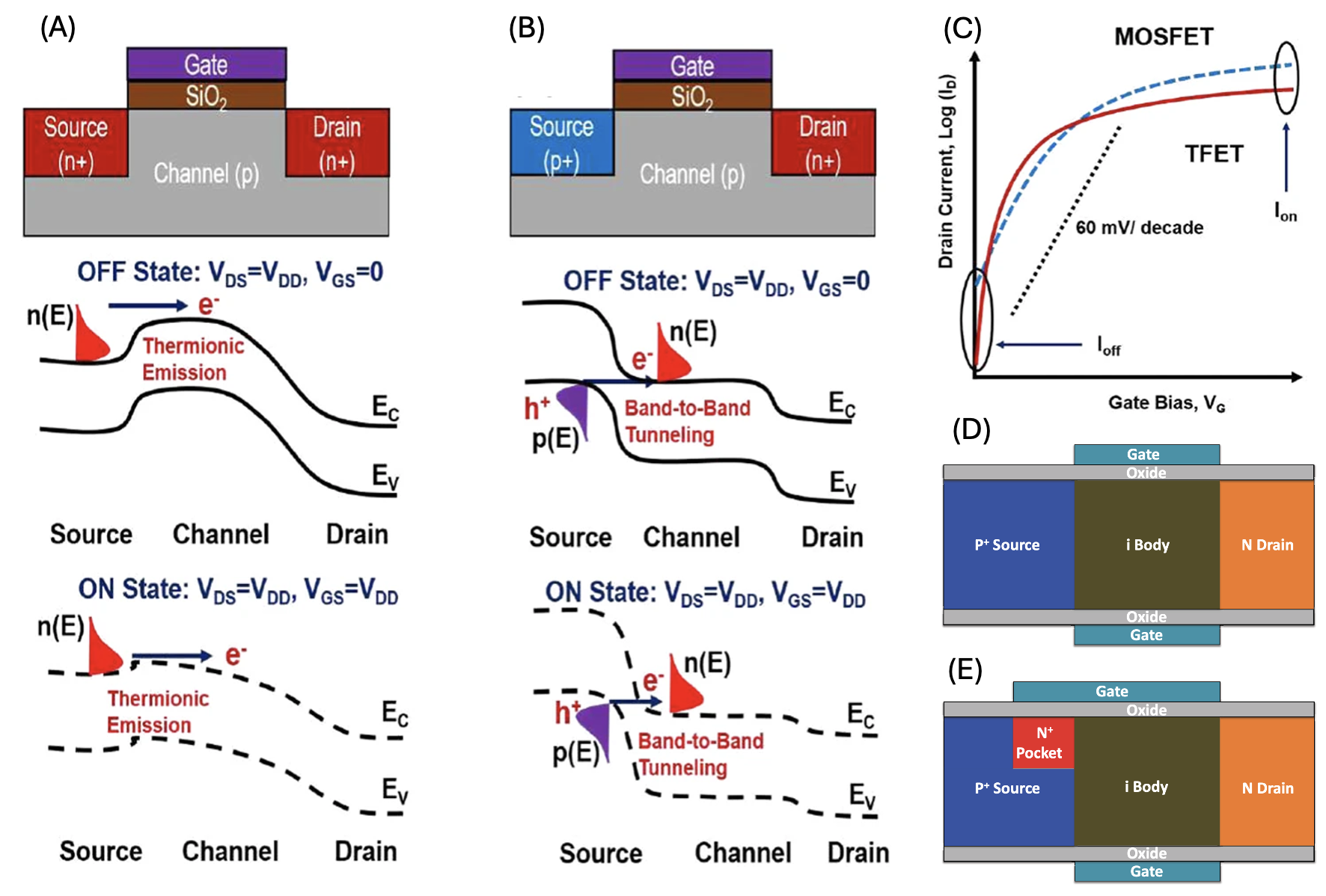}
    \caption{The energy band diagram of \textbf{(A)} MOSFET, \textbf{(B)} TFET for ON-state and OFF-state conditions, and \textbf{(C)} comparative transfer characteristics of well-designed MOSFET and TFET. Reproduced with permission from S. Kanungo et al., npj 2D Mater. Appl. 6, 83 (2022). Copyright 2022 Nature Publishing Group.\cite{kanungo_2d_2022}  Schematic of simulated devices, \textbf{(D)} Lateral InAs TFET and \textbf{(E)} Vertical InAs TFET with a heavily doped n+ pocket (halo) in the gate-source overlap region. Reproduced with permission from K. Ganapathi et al., Appl. Phys. Lett. 97, 033504 (2010). Copyright 2010 AIP Publishing.\cite{ganapathi2010analysis} }
    \label{fig:Figure1}
\end{figure*}

The fundamental nature of MOSFETs is such that their carrier modulation is not only challenged by the SCEs that accompany it, but it is also inherently limited in its ability to achieve an SS below the thermal limit.\cite{ionescu_tunnel_2011} In particular, the Boltzmann distribution of charge carriers encounters a thermal limit of 60 mV/dec at room temperature. In contrast to MOSFET devices, where an applied gate voltage reduces the energy barrier and enables thermionic carrier emission, TFETs operate at a specific applied gate voltage where the bands at the source and channel are modulated to tune the width of the source-channel barrier (Fig.1B). When the width is sufficiently reduced, BTBT can occur, which results in a notable enhancement in the device's switching speed between its off and on states.\cite{seabaugh_low-voltage_2010,kanungo_2d_2022} TFET devices employ this mechanism to facilitate electron tunnelling through the energy barrier from the conduction band minimum (CBM) to the valence band maximum (VBM).\cite{ionescu_tunnel_2011} It is also noteworthy that an energy barrier exists in both the on and off states, which constrains the on-state performance. This allows for the achievement of a sub-thermal SS (Fig.1C) and a significantly reduced power consumption in comparison to that of conventional FET devices.\cite{department_of_electronics_and_communication_engineering_fet_manav_rachna_international_university_faridabad_india_design_2018, kanungo_2d_2022} 

In terms of physical principles, BTBT is founded upon the quantum mechanical concept that a particle may traverse a potential energy barrier directly with a finite probability, contingent upon the barrier's width and height.\cite{chava_band--band_2023} The three primary parameters that exert the most significant influence on the tunneling probability are the carrier effective mass, bandgap, and screening length. The objective is to reduce these parameters in order to maximize the probability of tunneling. The Wentzel-Kramer-Brillouin (WKB) approximation represents the most general and widely utilized model for calculating the BTBT probability. The following equation defines the transmission probability: 
$$
    T_{\text{WKB}} \approx \exp \left(\frac{-4 \lambda E_g^{3/2} \sqrt{2m_t^*}}{3q \hbar (E_g + \Delta \phi)}\right)
$$
where $m_t$ is the effective mass of tunneling carrier, $E_g$ is the energy bandgap, $\hbar$ is the reduced plank constant, $q$ is the electronic charge, $\Delta \phi$ is the energy difference between VBM and CBM (energy window of tunneling), $\lambda$ is the screening length – spatial extent of energy band bending at tunneling junction.  The homogeneity of the material in homojunctions has been demonstrated to permit the WKB approximation to be precise in forecasting the existence of a solitary imaginary band that connects the real valence and conduction bands.\cite{ajoy_band_2012} This subsequently represents the dominant tunneling pathway. However, this model is only applicable to devices based on a single material (homojunctions) as it tends to overestimate the tunneling current for devices based on more than a single material (heterojunctions).\cite{kanungo_2d_2022} At the interface of heterojunctions, a discontinuity is observed in the imaginary wave vectors obtained from the complex band structures of the constituent materials.\cite{ajoy_band_2012}  In light of these considerations, heterojunctions are more accurately predicted by models such as the Kane model or others.\cite{salazar_predictive_2015, mazurak_wkb_2012, luisier_simulation_2010} It is of paramount importance to gain a precise understanding of the distinction between homojunctions and heterojunctions, as this affects the device's capacity to attain specific parameters. For example, when homojunctions exhibit disparate doping levels, it precludes the formation of a steep band profile, consequently broadening the width of the tunneling barrier.\cite{chava_band--band_2023} In contrast, the nature of heterojunctions allows for a reduction in the tunneling distance and screening length, thereby enhancing the transmission probability. Moreover, heterojunction devices often display an elevated BTBT current due to the diminished distance between the conduction and valence bands in comparison with homojunction structures. In particular, a reduction in field strength is sufficient for the generation of high currents.

TFETs are designed to operate in accordance with an applied gate voltage, which modulates the width of the tunneling barrier. The gate voltage can only control the width of the tunneling junction barrier by increasing the channel inversion, which represents a form of indirect modulation of the tunneling barrier. The conventional TFET configuration comprises a single gate, isolated by a dielectric material, mounted over a channel situated between the source and drain electrodes.\cite{kumar_kumawat_tunnel_2023} Prior research has indicated that double-gated structures demonstrate superior performance compared to single-gated TFETs. This is attributed to their capacity to mitigate ambipolar behavior and enhance the current within the device. A double-gated structure is precisely as its name suggests: an additional gate is placed parallel and opposite to the single gate, separated by dielectric layers. The use of heterojunctions of this kind can facilitate enhanced gate control due to the employment of a variety of gate materials with corresponding metal work functions. Furthermore, the double-gate structure has been shown to exhibit enhanced electrostatic control, a higher on/off current ratio, a higher on-current, and a lower off-current in TFET devices.\cite{usha2019compact} 

In terms of their architectural specifications, TFET devices are further classified as either horizontal or vertical, which indicates the direction of tunneling within the device. In contrast to the lateral carrier transport observed in horizontal TFET devices, BTBT in vertical TFETs can occur at an angle perpendicular to the gate oxide and channel interface.\cite{zhong-fang_han_simulation_2011} The differentiation between these devices is based on the distinction between their respective mechanisms for transitioning between the off and on states. In a lateral device (Fig.1D), when the gate voltage exceeds  (V\textsubscript{T}), the tunneling barrier width becomes sufficiently thin for BTBT to occur, resulting in the overlap of the conduction and valence bands.\cite{marin_lateral_2020} These conditions permit the occurrence of a substantial tunneling current, which in turn allows for a larger on-current. In the off state, the gate voltage is less than V\textsubscript{T}, and BTBT is not permitted due to the tunneling barrier width exceeding the permitted thickness. In this state, although some leakage current does occur, it is not significant. In the off state of a vertical device, a thin barrier is maintained, yet the absence of band overlap precludes BTBT. This distinction enables the vertical TFET to achieve a more compact SS.\cite{han_analysis_2013} Vertical TFETs (Fig.1E) permit direct modulation of the barrier width and enhanced gate control of BTBT. This shift in orientation has a significant impact on device performance, with vertical heterojunctions demonstrating superior capabilities compared to lateral heterojunctions. The regulation of the tunneling current through the gate voltage has enabled the achievement of a lower SS, which has resulted in a reduction in both the off current and power consumption.\cite{han_analysis_2013} The materials utilized for heterojunctions are distinct for the source and channel to achieve the requisite abrupt interface for the narrowing of the tunnel barrier width. The following sections will provide a more detailed examination of the various proposed structures.

\section{2D Materials + Heterojunctions Suited for High-performing  TFET devices and Optimization of Low SS}
An ideal high-performing TFET device should exhibit the following key electrical characteristics: a small SS, low power consumption, high on-state current, and minimal leakage current.\cite{lu_tunnel_2014} In the design of TFETs, the objective is to attain all the essential electrical characteristics of a high-performance device while minimizing the three parameters that predominantly influence BTBT probability.  In addition to SCEs that arise in MOSFETs, the selection of materials represents a significant factor influencing the overall performance of the device. The bandgap of the material is an intrinsic property that can either facilitate or impede the SS. The effective masses in the valence and conduction bands of the source and channel materials also exert an influence on the tunneling mass.\cite{kanungo_2d_2022} Moreover, the discovery of new semiconductor materials is crucial for the reduction of SCEs.\cite{chhowalla_two-dimensional_2016} In this regard, 2D materials have emerged as a promising avenue for TFET applications, given their high density of states, narrow thickness, and the absence of dangling bonds at the surface.\cite{lu_performance_2012} These attributes provide an excellent foundation for high electrostatic control. It is important to note, however, that the results have also demonstrated a tendency for high leakage current and low on-state current. This highlights the necessity for a comprehensive approach to achieve the optimal characteristics of TFETs through the optimization of material properties.

\subsection{Group III-V Materials}
Silicon (Si) is the material most commonly utilized in the contemporary semiconductor industry. The integration of TFET devices with Si allows for compatibility with existing fabrication processes and Si-based circuits, thereby facilitating the integration of new technology into existing infrastructure. Si, in particular, exhibits characteristics that are conducive to the development of TFETs. The indirect bandgap necessitates thermal activation for electron transitions, thereby facilitating the regulated reduction of off-state leakage. This is due to the fact that the thermal energy present at room temperature is insufficient to overcome the energy barrier. Germanium(Ge)-based TFETs have been the subject of investigation due to the favorable characteristics of germanium, including a small bandgap and high compatibility with Si.\cite{kao_direct_2012} In the context of transistors, where minimal heat dissipation is crucial for energy efficiency, indirect bandgap materials are typically preferred for fabrication due to their favorable characteristics, which include the ability to withstand high temperatures without significant degradation.

Nevertheless, indirect bandgap semiconductors exhibit suboptimal electron transitions due to a change in momentum, which can result in reduced energy dissipation, depending on the band structure.\cite{campbell_fabrication_2008, kao_direct_2012} In applications where the objective is to enhance energy efficiency by reducing thermal output, indirect bandgap semiconductors, such as Si, are frequently the preferred materials. In contrast, direct bandgap semiconductors demonstrate enhanced electronic transitions,\cite{kao_direct_2012} although this efficiency can result in elevated heat generation, which may be a disadvantage in certain applications. This renders them less suitable for applications involving traditional transistors. Nevertheless, they have been demonstrated to exhibit high efficiency with regard to BTBT in comparison to indirect bandgap materials. The direct bandgap enables direct electron tunneling with reduced energy requirements, resulting in a diminished SS, diminished off-state leakage current, and augmented energy efficiency.\cite{zhu_low-power_2013} Consequently, direct bandgap materials are preferred in the fabrication of TFETs due to the BTBT carrier injection method they employ.  

Group III-V materials, including indium arsenide (InAs) and indium gallium arsenide (InGaAs), have emerged as pivotal compounds in the enhancement of TFET performance due to their distinctive semiconductor properties (Fig.2 A-D).\cite{convertino_iiiv_2018, dewey_iiix2013v_2012} InAs, with its low power band gap of 0.35 eV, which is significantly lower than that of Si (1.12 eV), facilitates increased drain current through direct tunneling and is essential for achieving high on/off switching ratios in TFETs.\cite{rajan_performance_2022} This quality renders it an especially attractive option for TFET applications, establishing it as a prevalent choice for Group III-V TFET devices and a frequently featured material in relevant academic literature. The high electron mobility of InAs, which is several orders of magnitude greater than that of Si, is a significant contributing factor to this enhanced performance. Recent research has investigated the potential of InAs, InGaAs/GaAsSb, and InAs/GaSb heterostructure devices (Fig.2 E-I), with promising results.\cite{convertino_iiiv_2018,rajan_performance_2022,ganapathi_analysis_2010,lu_performance_2012, takagi_iii-vge-based_2017, moselund_inassi_2012} These studies, including one by Dutta et al., have exploited these properties to achieve subthreshold swings as low as 61.2 mV/dec and on/off ratios up to $7.13 \times 10^4$ in InAs-based double gate TFETs (see Table I. for performance parameters). These findings suggest that InAs offers substantial improvements in both on-state and off-state performance, making it a strong candidate for low-power, high-performance applications.

\begin{figure*}
    \centering
\includegraphics[width=1\linewidth]{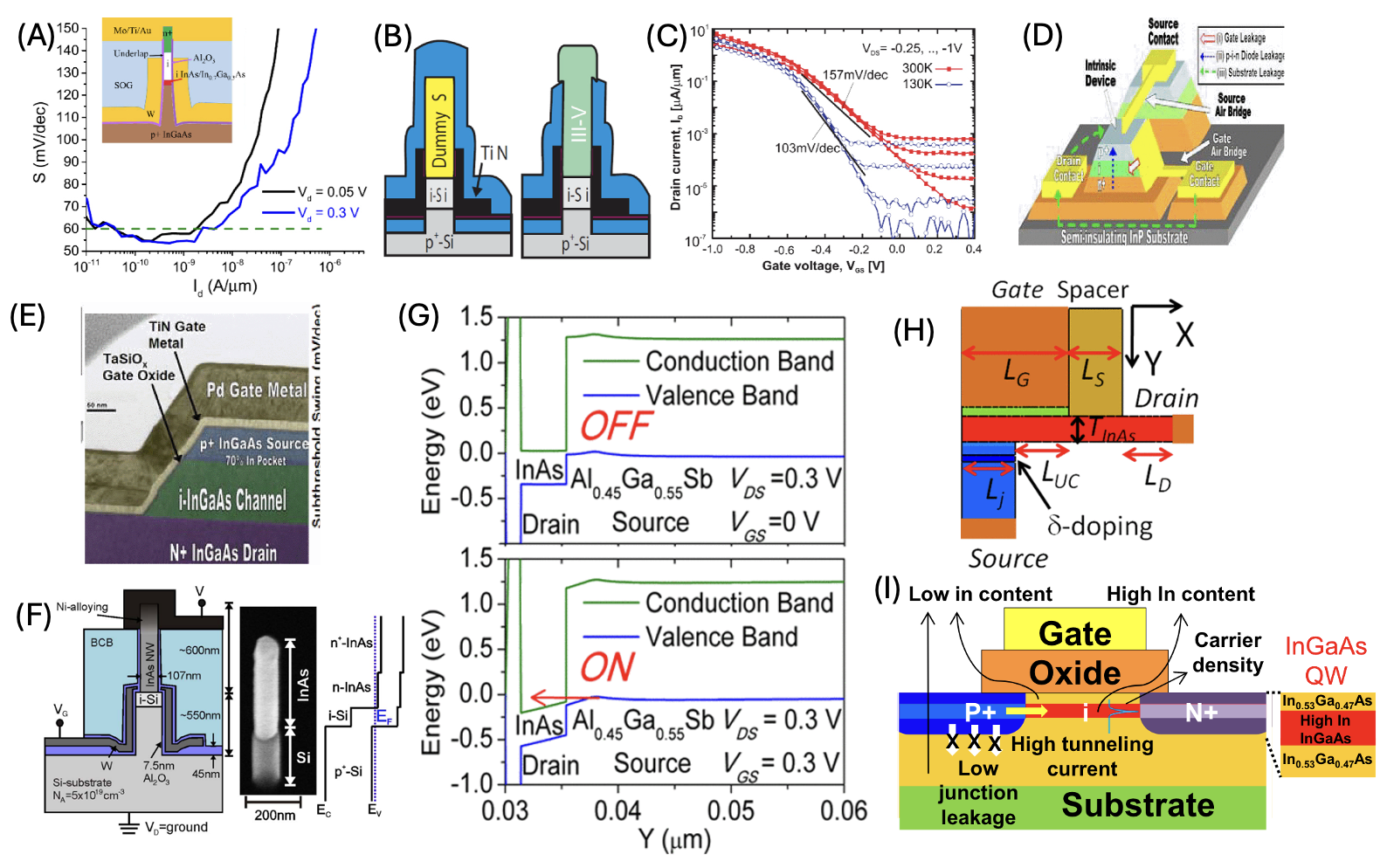}
    \caption{
    \textbf{(A)} SS versus I\textsubscript{DS} for a TFET fabricated using etched InGaAs/InAs heterostructure, demonstrating subthermal transport over two decades of current. \textbf{(B)} Example of template-assisted selective epitaxy (TASE) of a TFET heterostructure in a vertical nanowire. \textbf{(C)} Transfer characteristics of such a device. Reproduced with permission from C. Convertino et al., J. Phys.: Condens. Matter 30, 264005 (2018). Copyright 2018 IOP Publishing.\cite{convertino_iiiv_2018} \textbf{(D)} Schematic of the InGaAs Heterojunction TFET with a 5$\mu$m thick body and single gate. Reproduced with permission from G. Dewey et al., 2012 Symposium on VLSI Technology (VLSIT), 45–46 (2012). Copyright 2012 IEEE.\cite{dewey_iiix2013v_2012}  \textbf{(E)} TEM micrograph of InGaAs Heterojunction TFET showing the 4 nm ALD TaSiOx gate dielectric and the TiN/Pd metal gate. \textbf{(F)} Proposed InGaAs quantum well (QW) structure. Reproduced with permission from S. Takagi et al., 2017 Fifth Berkeley Symposium on Energy Efficient Electronic Systems \& Steep Transistors Workshop (E3S), 1–3 (2017). Copyright 2017 IEEE.\cite{takagi_iii-vge-based_2017} \textbf{(G)} On top, energy-band diagrams in the OFF state: V\textsubscript{DS} = 0.3 V and V\textsubscript{GS} = 0 V. On bottom, energy-band diagrams in the ON state: V\textsubscript{DS} = V\textsubscript{GS} = 0.3 V. \textbf{(H)} Two-dimensional cross section of the simulated AlGaSb/InAs staggered-gap n-channel TFET device structure. Reproduced with permission from Y. Lu et al., IEEE Electron Device Lett. 33, 655–657 (2012). Copyright 2012 IEEE.\cite{lu_performance_2012} \textbf{(I)} Schematic of the InAs Nanowire (NW) TFET and the SEM image showing a heterojunction NW after DRIE of Si. Ozone cleaning followed by HF treatments causes shrinking of the InAs compared with Si. To the right is a schematic of the energy band-edge diagram. Reproduced with permission from K. E. Moselund et al., IEEE Electron Device Lett. 33, 1453–1455 (2012). Copyright 2012 IEEE. \cite{moselund_inassi_2012}
}
    \label{fig:Figure2}
\end{figure*}

\begin{table}[ht]
    \centering
    \caption{Summary of key performance metrics of various TFET designs, including their on-current, subthreshold swing, and threshold voltage. Reproduced with permission from U. Dutta et al., Int. J. Mod. Educ. Comput. Sci. 10, 65–73 (2018). Copyright 2018 MECS Publisher.\cite{department_of_electronics_and_communication_engineering_fet_manav_rachna_international_university_faridabad_india_design_2018}}
    \includegraphics[width=0.4\textwidth]{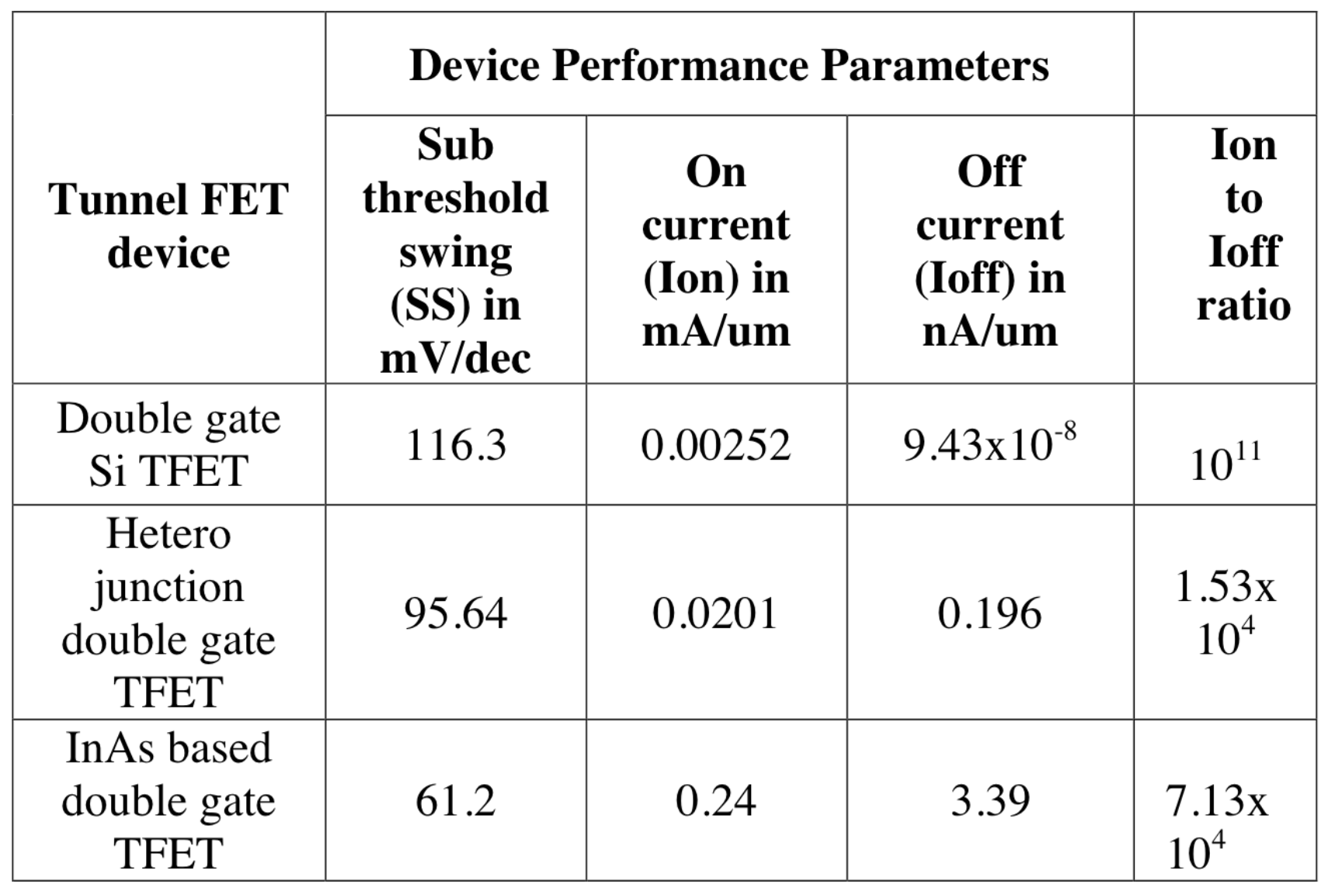}
    \label{tab:Table1}
\end{table}

The research conducted by Takagi et al.\cite{takagi_iii-vge-based_2017} explored an InGaAs/GaAsSb heterostructure (Fig.2 F), achieving an even higher on/off ratio of $10^9$ and a saturation speed of approximately 30 mV/dec.\cite{takagi2017steep} The selection of materials was deliberate, with the objective of targeting small and direct band gaps to enhance TFET on currents. Furthermore, they proposed a quantum well device with a Zn-diffused source region, which not only enhances the on current but also mitigates the off current due to the thin quantum well design, thereby attaining high on/off ratios at room temperature. The comprehensive review by Kumawat et al.\cite{kumar_kumawat_tunnel_2023} corroborates these findings, thereby reinforcing the notion that III-V compound semiconductors are the optimal choice for the source and drain in heterojunction TFETs.\cite{kumar_kumawat_tunnel_2023} The employment of materials with diminished direct band gaps has been evidenced to augment device functionality, elevating the on-current and mitigating the off-current. This, in turn, results in enhanced outcomes for leakage current and SS. Convertino et al.\cite{convertino_iiiv_2018} have demonstrated the versatility of III-V heterostructures through their exploration of InAs/GaSb, InAs/Si, and InGaAs/GaAsSb TFET structures.\cite{convertino_iiiv_2018} The findings indicate that while InAs/GaSb nTFETs encounter performance issues related to depletion and gate stack optimization, InAs/Si pTFETs demonstrate promising outcomes with an average SS of approximately 70 mV/dec. Moreover, the InGaAs/GaAsSb system has been put forth as a means of accommodating both p- and n-channel devices, thereby offering a potential avenue for the development of complementary TFET technologies.

The encouraging outcomes of III-V heterostructures in TFET applications (Table I) underscore the significance of continued research and development in this field. By continuing to leverage the properties of these materials, such as high electron mobility and direct tunneling facilitated by narrow band gaps, enhancements of TFET performance can facilitate the development of low-power and higher-efficiency electronic devices. 

\subsection{Transition Metal Dichalcogenides (TMDs)}
2D TMDs are distinguished by their ultra-thin body, which has the effect of enhancing gate control and thus reducing SCEs.\cite{resta_devices_2019} Due to their stackable nature and tunable thickness, TMDs can be precisely configured to exhibit a desired band structure and electronic properties, including the bandgap.\cite{ilatikhameh2015tunnel} Multilayer TMDs typically exhibit indirect band gaps, rendering monolayer TMDs (which possess direct band gaps) more conducive to experimentation. TMD materials, such as molybdenum disulfide (MoS\textsubscript{2}) and tungsten disulfide (WS\textsubscript{2}), possess a direct bandgap, which is optimal for the tunneling carrier injection mechanism within a TFET device. In addition to these properties, TMDs are known to offer a high on/off current ratio, low SS, and high carrier mobility. However, TMDs are known to possess band gaps exceeding 1 eV, which precludes their use in TFETs for logic applications due to the inability to achieve the requisite drive current.\cite{zhang_two-dimensional_2014} While trade-offs exist with regard to device capabilities in the context of large or small bandgap, depending on the application, the use of TMDs remains a highly advantageous proposition. Smaller band gaps, exemplified by III-V materials as previously discussed, are preferred for high-speed switching applications and low operating voltages.\cite{zhang2012analysis} In contrast, the use of materials with substantial band gaps, such as TMDs, frequently exhibits greater advantages in the context of augmented thermal and electrical stability, hybrid TFET designs, and high-reliability applications.

\begin{figure*}
    \centering    \includegraphics[width=1\linewidth]{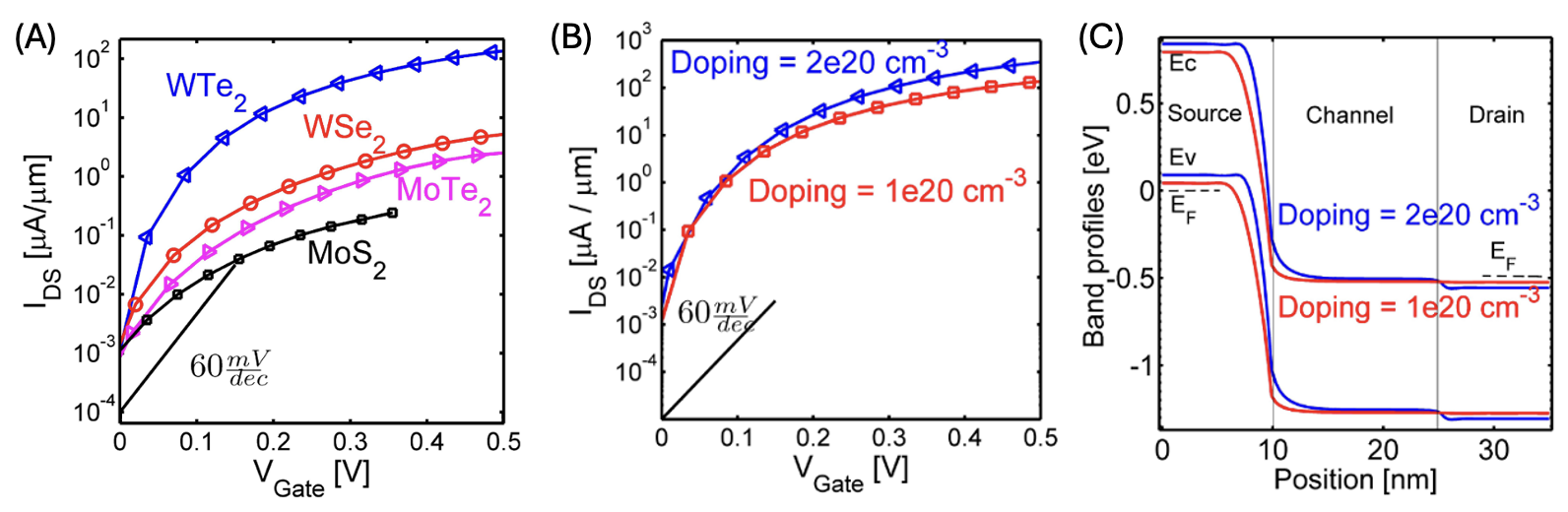}
    \caption{
    High performance of TMD materials. \textbf{(A)} Transfer characteristics of TMD TFETs with I\textsubscript{OFF} = 1 nA/$\mu$m.\textbf{(B)} Transfer characteristics and \textbf{(C)} band diagrams of WTe\textsubscript{2} with doping levels of 1 $\times$ 10\textsuperscript{20} and 2 $\times$ 10\textsuperscript{20} cm\textsuperscript{-3}. Reproduced with permission from H. Ilatikhameneh et al., IEEE J. Explor. Solid-State Comput. Devices Circuits 1, 12–18 (2015). Copyright 2015 IEEE.\cite{ilatikhameh2015tunnel}
    }
    \label{fig:Figure3}
\end{figure*}

In a study conducted by Joshi et al., the MoTe\textsubscript{2} TFET was proposed as a device for visible light detection and photosensor applications.\cite{joshi_photosensor_2020} The device configuration employed was a DMG-TFET, wherein MoTe\textsubscript{2} was utilized as the channel material, exhibiting a thickness of 0.65 nm and a channel length of 100 nm.\cite{joshi_photosensor_2020} The employment of this TMD material was found to result in a low energy band gap (0.8-0.11 eV), which yielded high on-current and high sensitivity for the photosensor (high transmission in the visible range) in comparison to other TMDs. Furthermore, WTe\textsubscript{2} has been identified as a promising candidate for TFET due to its superior on-current characteristics and reduced DIBL effects (Fig.3A). This is due to the fact that, in comparison with MoTe\textsubscript{2}, WTe\textsubscript{2} exhibits smaller in-plane dielectric constants, which serve to reduce electric field penetration from the drain and suppress SCEs. Notwithstanding the larger bandgap of WTe\textsubscript{2}, the on-current is situated in closer proximity to the threshold voltage. This phenomenon can be attributed to the lower dielectric constants of WTe\textsubscript{2}, which serve to enhance the on-current. For this TFET, the minimum achievable current is exceedingly low, remaining below 1 nA/$\mu$m even with higher drain doping levels (Fig.3 B-C).  In light of these observations, it can be concluded that WTe\textsubscript{2} is a promising candidate for TFET applications, as evidenced by its superior performance in various metrics, including on-current, SS, DIBL, and energy-delay product. These findings underscore the importance of considering not only thin-channel materials but also the optimal combination of bandgap, effective mass, and doping concentrations to achieve high-performance TFETs. It is evident that the mere thinning of materials is insufficient for optimizing device performance; attention must also be paid to the design choices surrounding the epitaxial layer thickness, body thickness, and doping levels.

\subsection{Black Phosphorous (BP)}
BP has emerged as a leading contender for next-generation TFETs,\cite{lu2017performance} largely due to its distinctive properties as a 2D material. Its high electron mobility is instrumental in facilitating rapid charge transport, which is crucial for high-speed electronic devices.\cite{li_structures_2015} It is noteworthy that the electron mobility in monolayer BP is high, reaching approximately 10,000 cm\textsuperscript{2}/V·s. In contrast, the electron mobility is observed to decrease to approximately 1,000 cm\textsuperscript{2}/V·s in multilayered structures. This variation in mobility with thickness is of critical importance, as it allows for the precise engineering of the transistor's electrical properties.\cite{he_tuning_2019} Moreover, the direct bandgap of BP is dependent on the layer and can be precisely adjusted, which is crucial for modulating the transistor's operating wavelength and for applications in optoelectronics. The electrical conductivity of BP can be effectively turned on and off by controlling the thickness, achieving on/off ratios of up to 10\textsuperscript{4}\textasciitilde10\textsuperscript{5}, which is a significant advantage for digital switches where distinct states of current flow are essential. In addition to these electronic properties, the ease with which its thickness can be manipulated represents a considerable advantage.\cite{lu_modeling_2017}

An array of BP-based devices has demonstrated the potential for applications that require low-power and efficient switching, as evidenced by experimental results.\cite{nazir_energy-efficient_2020} Among these, a BP TFET with modulated thickness is worthy of particular note for its suitability for low-power applications.\cite{kim2020thickness} Kim et al. have developed two BP natural heterojunction (NHJ)-TFETs at V\textsubscript{D} $\leq$ 0.7V: device 1 has a bottom-gate dielectric of 285-nm SiO\textsubscript{2} and top-gate dielectric of 10-nm hBN, and device 2 has a bottom-gate dielectric of 3-nm hBN and top-gate dielectric of 5-nm hBN. The BP device 1 exhibits a low SS of 23.7 mV/dec (Fig.4A), with an averaged value of 4–5 decades.Furthermore, the device exhibits a considerable on-current, the measured drain current (I\textsubscript{D}) versus V\textsubscript{TG} showing  I\textsubscript{60}= 0.65 $\mu$A/$\mu$m. A noteworthy advancement has been the demonstration of BP TFETs with bilayer hBN tunnel barriers at the drain contact as highly promising switching devices. The bilayer hBN construction resulted in superior device efficiency, as indicated by an I\textsubscript{60} of 0.65 $\mu$A/$\mu$m and an SS of <60 mV/dec (averaged for four decades) at 300 K.\cite{kim2020monolayer} The bilayer hBN exhibited a markedly reduced V\textsubscript{T} change of 0.1 V, in stark contrast to the typical 0.7 V change observed in conventional MOSFETs. Moreover, the pivotal function of hBN was investigated in a BP device that demonstrated remarkable performance, with a record-high I\textsubscript{60} of 19.5 $\mu$A/$\mu$m at a drain voltage (V\textsubscript{D}) of -0.7 V (p-type), and an SS = 37.6 mV/dec (averaged for four decades) at 300 K. In contrast, Wu et al. presented an alternative complementary BP TFET design that exhibited disparate characteristics, as shown in Fig. 4B.\cite{wu_complementary_2019} The findings indicated that the minimum SS was 178 mV/dec, which exceeded the Boltzmann limit significantly. This was attributed to the thickness of the BP flake used, which ranged from 8 to 13 nm. By precisely adjusting the channel thickness and reducing the equivalent oxide thickness (ETO) to approximately 0.5 nm, the researchers were able to significantly enhance the performance, resulting in an on-current of 800 $\mu$A/$\mu$m and an SS of 12 mV/dec.

The integrity of a semiconductor's crystal lattice is of paramount importance with regard to the electronic properties exhibited by the material. Similarly, the electronic characteristics of BP are closely related to its structural purity,\cite{li_structures_2015} thereby reinforcing the importance of high-quality material synthesis for advanced electronic applications such as TFETs. Consequently, the achievement of single-crystalline 2D materials represents a crucial objective within the domain of semiconductor device fabrication.  By modulating the thickness of the material, it is possible to tailor BP in order to reduce the incidence of interface defects.\cite{li2015structures, he_tuning_2019, kim2020thickness} Defects of this nature, including those of a lattice mismatch at surfaces, not only impair the intrinsic properties of the material in question but can also introduce trap levels that impede the flow of charge. In an ideal semiconductor, the absence of impurities and defects would result in the absence of electronic states within the band gap. However, the presence of impurities, such as transition metals, often results in the formation of deep levels, which are energy states situated at a considerable distance from the band-gap edges. Such defects can function as traps for charge carriers, thereby impeding the conductivity of the device.\cite{li2015structures}

\begin{figure*}
    \centering
\includegraphics[width=0.9\linewidth]{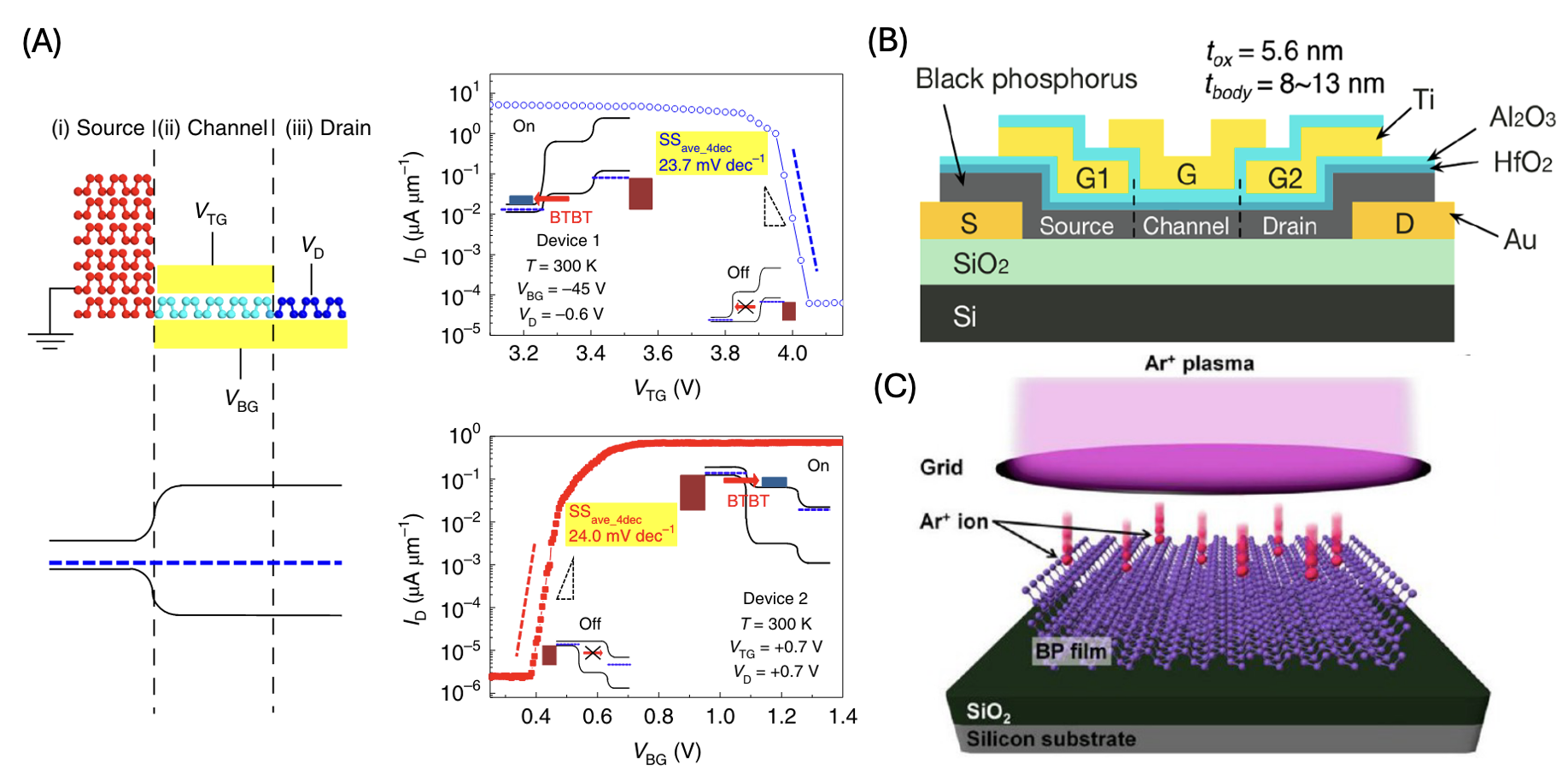}
    \caption{\textbf{(A)} BP natural heterojunction (NhJ)-TFET schematic structure and BP band diagram in the (i) source, (ii) channel and (iii) drain. Reproduced with permission from S. Kim et al., Nat. Nanotechnol. 15, 203–206 (2020). Copyright 2020 Nature Publishing Group.\cite{kim2020thickness} \textbf{(B)} Schematic of the BP reconfigurable electrostatically doped (RED) TFET. Reproduced with permission from P. Wu et al., ACS Nano 13, 377–385 (2019). Copyright 2019 American Chemical Society.\cite{wu_complementary_2019} \textbf{(C)} Schematic representation of the Ar+ plasma treatment process to BP for  defect-tailoring. Reproduced with permission from D.-H. Kang et al., ACS Photonics 4, 1822–1830 (2017). Copyright 2017 American Chemical Society.\cite{kang_self-assembled_2017}}
    \label{fig:Figure4}
\end{figure*}

Defects in BP can be classified according to three criteria: the nature of the bond, the structural distortions they induce, and the manner in which the bonds are broken. Such structural deformations, including vacancies in the crystal lattice, have the potential to significantly alter the band gap energy of the material. For instance, a modified BP with a divacancy of the P1-P2 type (where P1 and P2 indicate the positions of two phosphorus atoms) has the potential to undergo a transition from its characteristic direct band gap to an indirect one, with a value of 1.02 eV.\cite{li_structures_2015} This transition is particularly disadvantageous for TFETs design, as these devices are optimized for BTBT, which is more effective with direct-bandgap semiconductors. The probability of tunneling is higher with direct BTBT due to the alignment of the valence and conduction bands in k-space, which facilitates a direct recombination of electrons and holes.\cite{kanungo_2d_2022} To optimize BP for TFETs, it is essential to implement high-purity fabrication and effective defect management. Techniques such as optimized chemical vapor deposition\cite{li_electrochemical_2020} or annealing\cite{li_high-performance_2016} can be employed to mitigate defect-induced alterations to the bandgap. It is noteworthy that controlled defect engineering\cite{he_tuning_2019} could potentially be employed to precisely adjust BP's electronic properties for specific device functions or to develop novel semiconductor devices that operate on disparate principles, such as resonant tunneling. An example from Kang et al. is using an argon plasma treatment process to BP for defect-tailoring, shown in Fig.4C.\cite{kang_self-assembled_2017} Despite the inherent challenges, defects can be harnessed to enhance TFET functionality when managed strategically.

\section{Multi-Scale Simulations}
MS simulations are an indispensable tool for advancing TFETs, as they provide insights that are not readily accessible through experimental techniques alone.\cite{fiori_multiscale_2013} They facilitate a comprehensive assessment of material properties, device physics, and operational characteristics at the nanoscale,\cite{horstemeyer_multiscale_2010} which are crucial for optimizing the performance and reliability of these devices. By employing MS simulations, a wide range of materials (including homojunction\cite{fiori_multiscale_2013} and heterojunction materials\cite{huang_multiscale_2017} and device geometries) can be explored to identify optimal TFET designs, thus obviating the monetary costs and time consumption associated with experimental testing of physical prototypes.\cite{huang_quantum_2016} As device dimensions continue to decrease, traditional fabrication techniques and materials may introduce additional opportunities for error, necessitating the development of new approaches to ensure the reliability of the final product. MS simulations can thus anticipate these issues and allow for adjustments in either the design or the materials used to avoid them.\cite{polat_modified_2021} Similarly, the electrical characteristics of TFETs can be simulated under a variety of conditions to provide key performance metrics, including on/off ratios, SS, and overall efficiency. A comparison of these characteristics with the requirements set out in the International Technology Roadmap for Semiconductors (ITRS) and with those of other devices is essential for the advancement of TFET technology. 

 The domain of MS simulations is anchored in three main categories: heterojunctions, homojunctions, and the emergent class of 2D materials. In this discussion, we will examine the complexities of simulation effectiveness across these domains, with a particular focus on the intricacies of TMD heterojunction simulations, the dynamics of BP homojunctions, and 2D materials such as arsenene (As), antimonene (Sb), and monolayer BP. In this section, we present a synthesis of findings from the existing literature to provide a comprehensive overview. In addition, we evaluate the range of their applications in different sectors and assess their significance in relation to the goal of advancing functionality and innovation within TFET technologies. The efficacy of MS simulations in this context for MOSFET development and material exploration underscores the pivotal role of MS simulations in propelling the advancement of TFET technology. 

\subsection{Framework of Quantum Simulations}
\begin{figure*}
    \centering    \includegraphics[width=1\linewidth]{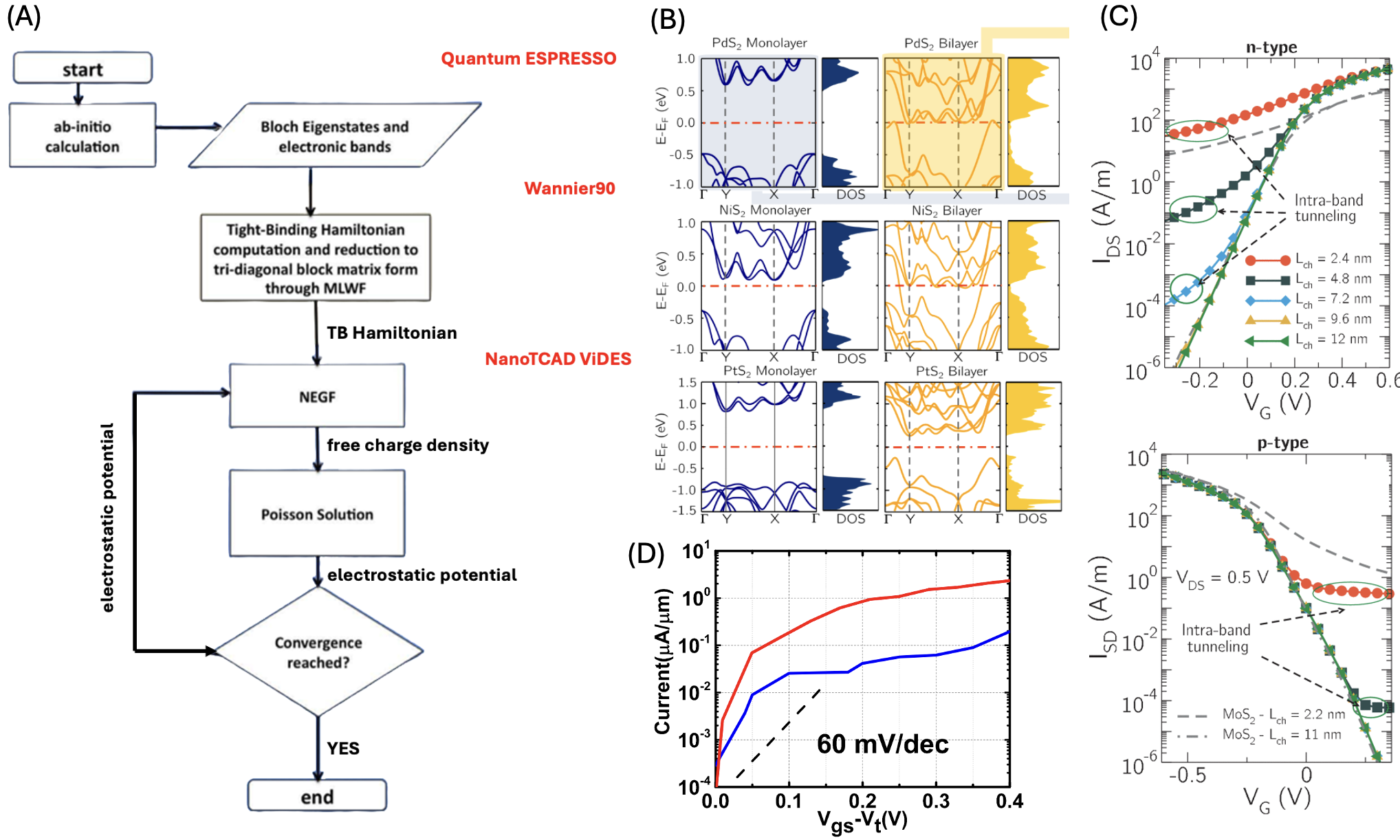}
    \caption{
    \textbf{(A)} Flowchart of an open-source multi-scale framework for simulation of nano-scale devices. Reproduced with permission from S. Bruzzone et al., IEEE Trans. Electron Devices 61, 48–53 (2014). Copyright 2014 IEEE.\cite{bruzzone_open-source_2014} 
\textbf{(B)} Electronic band structure along a symmetric path in the Brillouin zone and DOS computed with DFT for monolayer and bilayer PdS\textsubscript{2}, NiS\textsubscript{2}, and PtS\textsubscript{2}. Reproduced with permission from E. G. Marin et al., ACS Nano 14, 1982–1989 (2020). Copyright 2020 American Chemical Society.\cite{marin_lateral_2020}
\textbf{(C)} Simulated transfer characteristics of the n-type (left) and p-type (right) InSe FETs for V\textsubscript{ds} = 0.5 V and several channel lengths with t\textsubscript{ox} = 0.5 nm and L\textsubscript{ch} = L\textsubscript{g}. MoS\textsubscript{2} FETs characteristics (dashed lines) are included for comparison purposes. Reproduced with permission from E. G. Marin et al., IEEE Electron Device Lett. 39, 626–629 (2018). Copyright 2018 IEEE.\cite{marin_first-principles_2018} 
\textbf{(D)} Simulated I\textsubscript{DS}-V\textsubscript{GS} curves for BP VTFET, 3.3 nm long device with S/D doping (blue) and undoped D (red). Reproduced with permission from S.-C. Lu et al., 2017 International Conference on Simulation of Semiconductor Processes and Devices (SISPAD), 345–348 (2017). Copyright 2017 IEEE.\cite{lu_modeling_2017} 
    }
    \label{fig:Figure5}
\end{figure*}

The application of an ab-initio quantum framework that integrates DFT, maximally localized Wannier functions (MLWF), and non-equilibrium Green functions (NEGF) facilitates the accurate estimation of transport properties and comprehensive device performance. This is a critical element in the process of guiding experimental design and device optimization. The steps required to perform these simulations and predict the transport properties of transistor devices, and its framework is outlined in Fig. 5A.\cite{fiori_multiscale_2013, bruzzone_open-source_2014} In the first step, a DFT package (e.g., Quantum ESPRESSO) is used to perform a series of calculations that predict not only the electronic structure but also the thermodynamic properties, providing a comprehensive picture of the intrinsic properties of the channel material. The Hamiltonian of the channel material is then transformed from a Bloch basis of extended eigenstates to a basis of MLWFs by Wannier90. These Wannier functions defined in terms of Bloch eigenstates are subjected to unitary transformations over reciprocal space, yielding generalized Wannier functions.  While these functions are not inherently localized, localization is enforced by solving a function minimization problem. This process yields MLWFs that provide an effective tight-binding (TB) Hamiltonian for the electronic bands near the fundamental gap and facilitate efficient band interpolation. Elements within the TB Hamiltonians act as fitting parameters, allowing the calculation of carrier transport and charge distribution corresponding to each atom. Finally, the Hamiltonian with this basis of calculated MLWFs allows the calculation of current and transmission coefficients and properties such as electron and hole concentrations by the NEGF method (e.g., using NanoTCAD ViDES via a self-consistent NEGF and Poisson solver).

\subsection{Applications of MS Simulations}
\subsubsection{Graphene}

Despite its lack of intrinsic band gap, graphene has exceptional electrical properties that make it an interesting candidate for electronics.  Fiori and Iannaccone et al.\cite{fiori_multiscale_2013} describe graphene-based transistors through MS modeling, presenting graphene nanoribbon (GNR) transistors,\cite{lone2021review} graphene bilayer FETs,\cite{szafranek2012current} and hexagonal boron-carbon-nitride(hBCN)/graphene heterostructures. The detailed MS approach helps to overcome the limitations of graphene's zero bandgap through strategies such as bandgap engineering by chemical functionalization or the use of graphene in complex heterostructures. For GNR FETs, the simulations predict large I\textsubscript{on}/I\textsubscript{off} ratios for narrow devices, with performance strongly influenced by edge disorder and chemical modifications. Bilayer graphene FETs exhibit a bandgap modifiable by an external electric field, which is exploited in TFETs to achieve low-power operation suitable for digital applications. In addition, hBCN can be innovatively used as a barrier material in graphene channels, effectively blocking band-to-band tunneling, leading to high I\textsubscript{on}/I\textsubscript{off} ratios and demonstrating the potential of 2D graphene in advanced electronics. By providing insights into quantum effects and the electrostatic properties of materials at different scales, these simulations are crucial for optimizing material synthesis and device architecture.

\begin{table*}[ht]
    \centering
    \caption{Figures of merit for different channel lengths of the LH FET and DG planar barristor using monolayer MoS\textsubscript{2} for HP and low-power (LP) applications. Reproduced with permission from D. Marian et al., Phys. Rev. Appl. 8, 054047 (2017). Copyright 2017 American Physical Society.\cite{marian_transistor_2017}}
    \includegraphics[width=1\textwidth]{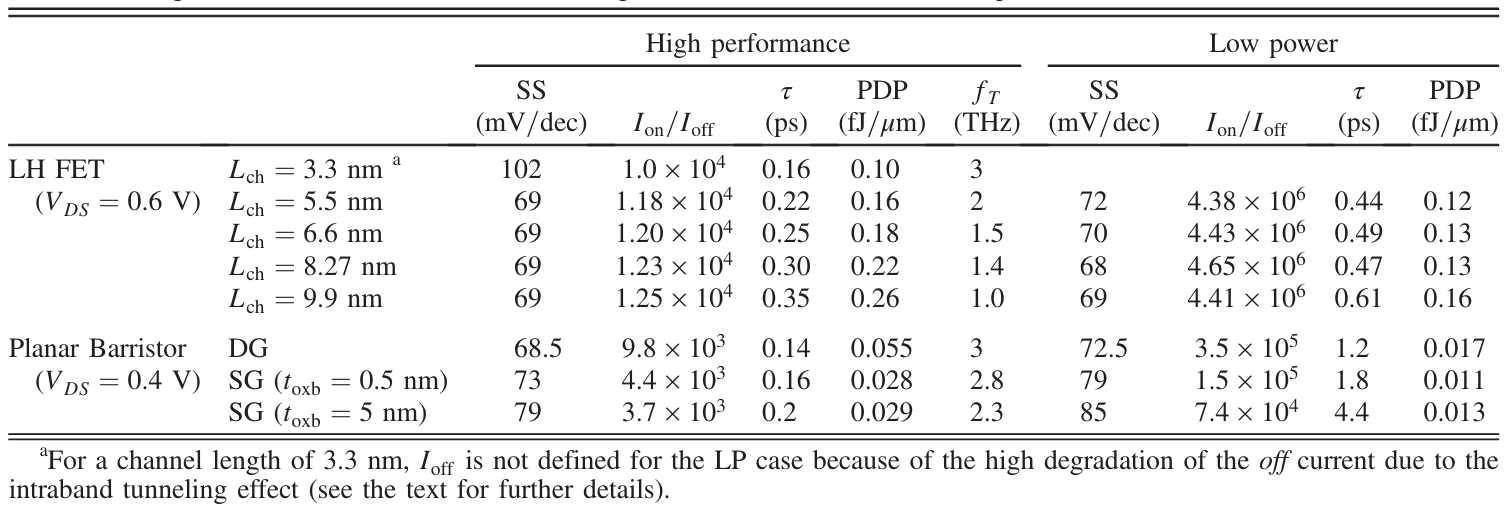}   
    \label{tab:Table2}
\end{table*}

 \subsubsection{TMDs}
MoS\textsubscript{2} TFET devices are also being investigated as a promising TMD for TFET applications, with numerous MS simulation studies supporting this research. Marian et al.\cite{marian_transistor_2017} contribute to this body of work by using MS simulations to introduce two advanced transistor concepts based on lateral heterostructures within a monolayer of MoS\textsubscript{2} that integrates adjacent metallic (1T) and semiconducting (2H) phases. These concepts are highly regarded for application in both high-performance and low-power devices. The paper discusses a lateral-heterostructure (LH) TFET with a semiconducting MoS\textsubscript{2} channel sandwiched between metallic MoS\textsubscript{2} regions, designed for superior electrostatic control and operating efficiency. A second concept is the planar barristor - a laterally gated Schottky diode - which effectively connects a metallic source to a semiconductor drain. The LH FETs feature near-ideal SS of 69-100 mV/dec over various channel lengths (Table II), providing excellent electrostatic control for high- performance applications. They also exhibit impressive I\textsubscript{on}/I\textsubscript{off} ratios that not only exceed 10\textsuperscript{4} for high-performance requirements, but also exceed 10\textsuperscript{6} for low-power requirements with channel lengths of at least 5.5 nm. In parallel, the planar barristor, especially in its double-gate configuration, exhibits SS values below 79 mV/dec, reflecting its gating efficiency. Its on/off ratio approaches 10\textsuperscript{4}, reinforcing its potential as a formidable competitor to conventional CMOS technology. These results, which encapsulate the devices' switching capabilities and mastery of off-state current leakage, suggest that MoS\textsubscript{2}-based lateral heterostructures hold great promise for the next generation of transistor technology, marking a step forward in the quest for devices that balance high performance with low-power consumption.

The transport properties of monolayer and bilayer configurations of PdS\textsubscript{2}, PtS\textsubscript{2}, and NiS\textsubscript{2} were also calculated (Fig.5B).\cite{marin_lateral_2020} The results from Marin et al.\cite{marin_lateral_2020} indicate that LH-FETs fabricated with NiS\textsubscript{2} do not meet the IDRS benchmarks due to its minimal bandgap and inherent ambipolar characteristics. However, LH-FETs fabricated with PdS\textsubscript{2} and PtS\textsubscript{2} meet the IRDS performance criteria, demonstrating their potential for integration into future high-performance digital applications. These results confirm the potential of 2D-based FETs beyond graphene to transcend the subthermal limit, thus inviting further research into 2D materials with more acute DOS and lower SS. In particular, noble TMDs have been instrumental in achieving subthermal SS in FETs at ambient conditions, mainly due to their distinct DOS properties.

The insights provided by these MS simulations help researchers and device designers by providing a predictive benchmark against which to measure and refine their fabrications. This predictive capability not only provides information on expected transfer characteristics, but also outlines the underlying physical principles that govern device behavior, which is critical to the design of advanced semiconductor devices. For example, InSe, with its high mobility and favorable bandgap of about 1.5 eV,\cite{hao2023bandgap} is emerging as an ideal candidate for the fabrication of ultrathin digital electronics. For n-type FETs, an SS of 65 mV/dec and an Ion/Ioff ratio greater than 2.7$\times$10\textsuperscript{4} were predicted for devices with a channel length of 7.2 nm.\cite{marin_first-principles_2018} These transfer characteristics from Marin et al., as seen in Fig.5C, indicate strong potential for high-performance applications, while also pointing to significant source-to-drain tunneling effects in shorter channels.\cite{marin_first-principles_2018} This is an important consideration for future device miniaturization. The p-type FETs exhibit less tunneling due to the larger effective hole mass, enabling robust performance at channel lengths greater than 7.2 nm. The improved performance of InSe FETs, despite their sensitivity to variations in oxide thickness, particularly in p-type devices, and their stability against gate length variations - with minimal performance degradation even at 30\% gate underlap - provide practical insights into the fabrication of consistent and reliable transistors. These InSe FETs not only outperform MoS\textsubscript{2} nFETs in terms of on-current and on-off ratio, but also exhibit greater robustness against intraband tunneling, a critical advantage over MoS\textsubscript{2} pFETs. Such transfer characteristics provide experimentalists with concrete performance benchmarks to aim for, thus influencing the trajectory of future experimental efforts.

\subsubsection{Black Phosphorous}
The BP-based TFET device is noteworthy for its potential in achieving ultra-scaled, low-power, and steep subthreshold logic devices due to the excellent electrostatic control afforded by 2D materials' narrow thickness. MS simulations employing a quantum simulation framework on BP TFET devices in diverse structural configurations (including heterojunction and homojunction arrangements) and varying chemical doping concentrations facilitate the optimization of device design and energy-delay metrics. The simulations indicate that tri-layer BP TFETs exhibit remarkably high on-currents in comparison with tri-layer WTe\textsubscript{2} DG TFETs.\cite{agarwal2018material} This performance is attributed to the superior material properties of tri-layer BP, namely a smaller effective mass and a larger transverse effective mass, which enhance the transmission probability and on-state current. Additionally, the bandgap of the tri-layer BP (E\textsubscript{g} = 0.76 eV) contributes to this effect. The device structure of this study also included an interfacial layer (IL) between the dielectric and two-dimensional material. This resulted in an increase of the on-current by three to four times its original value and served to mitigate the limiting effects of fringing fields at the source channel junction. Thus, the device performance can be significantly enhanced by effectively reducing the tunneling distance and shaping the potential distribution to be steeper at the junction.

Moreover, MS simulations provide a means of conducting energy-delay analysis, thereby enabling the assessment of a range of off-currents and supply voltages. Such evaluations demonstrate that tri-layer BP TFETs can maintain energy efficiency in comparison with monolayer BP FETs. The energy-delay product analysis demonstrated that for a target off-current (I\textsubscript{off}) of 10\textsuperscript{-5} $\mu$A/$\mu$m, the tri-layer BP TFETs exhibited superior delay and energy-delay product (EDP) characteristics across a range of supply voltages (V\textsubscript{DD}), particularly outperforming monolayer BP FETs at supply voltages below 0.5V.\cite{agarwal2018material} It is therefore evident that the capacity to undertake simulations that encompass a range of scales is of paramount importance to the advancement and innovation of TFET technologies. Such simulations provide a framework for predictive modeling and design, enabling the resolution of current challenges and the guidance of future advancements within this field. The study demonstrated the promising results of tri-layer BP TFETs using the proposed device design with IL, which exhibited superior performance compared to existing 2D FETs at lower supply voltages. Furthermore, the utility and efficiency of the MS simulations employed underscore the significance of advancing the functionality and innovation within TFET technologies.

Lu et al. also examined the utilization of BP for vertical TFET devices with asymmetric layer numbers for the top and bottom layers and an undoped drain by employing MS simulations.\cite{lu_modeling_2017} Moreover, the impact of varying the number of layers in the source material is examined, along with device performance with and without chemical doping. The results (as seen in Fig.5D) demonstrated that the SS and on/off current ratio for this device structure can be maintained below 10 mV/dec and beyond 10\textsuperscript{5}, respectively, when the channel length is reduced to 3 nm (Fig.5D). Even at a channel length of 3.3 nm, the device exhibits a relatively low SS of approximately 6 mV/dec, demonstrating a lesser degree of degradation in SS and on/off ratio in comparison with devices with conventional source/drain doping.\cite{lu_modeling_2017} The design exploits the layer-dependent properties of BP and its exceptional electrostatic control, which extends to the off-state, a crucial aspect for ultra-short channel TFETs aiming to minimize off-current while optimizing device performance. It was observed that for channel lengths below 10 nm, the use of an undoped drain can result in enhanced device performance, while an increase in on-current can be achieved by increasing the number of layers in the source. Furthermore, the on-current can be increased by another order of magnitude through the implementation of alterations in the channel orientation, specifically from zigzag to armchair.

\subsubsection{Newly Found 2D Materials}
Group-V materials, renowned for their uniquely buckled honeycomb configurations such as As-ene, Sb-ene, and bismuth-ene (Bi), have been identified as promising candidates for TFET design owing to their ambient stability and small effective masses, as outlined by Kanungo et al.\cite{kanungo_2d_2022} in their study on nanoscale TFETs. Amongst these, Bi is particularly distinguished for its minimal and direct energy bandgap. The crystal structure of Bi\textsubscript{2}Se\textsubscript{3} is rhombohedral with a nominal direct energy bandgap. Zhang et al.\cite{zhang_two-dimensional_2014} utilized ab-initio simulations to assess a Bi\textsubscript{2}Se\textsubscript{3} thin film, revealing a bandgap of 0.252 eV and positing its utility as a TFET channel material optimized for low-power logic applications. This material exhibited a subthermal SS over 4 orders of magnitude of 50 mV/dec at V\textsubscript{DS}=0.2V (Fig.6A) and a robust I\textsubscript{on}/I\textsubscript{off} ratio under minimal supply voltage, an advance reported by Zhang et al.\cite{zhang_two-dimensional_2014} Further, Li et al. extended ab-initio simulations to assess the wider spectrum of group-V materials—As, Sb, and Bi—and their deployment in 10 nm gate-length TFETs (Fig.6B).\cite{li_negative_2019} Monolayer bismuthine emerged with the highest on-state current, satisfying the ITRSbenchmarks for high-performance devices. Additionally, these group-V material-based TFETs showcased considerably reduced delay times and power dissipation, compared to ITRS standards. Among the hexagonal monolayer group V-ene contenders, the monolayer Bi TFET was identified by Li et al. as offering superior device performance in terms of on-state current, delay time, and power dissipation for high-performance applications.\cite{li_heterostructures_2016} Additionally, the monolayer Sb TFET demonstrated favorable performance, ranking closely behind the Bi TFET. These investigative simulations extend to the domain of MOSFET performance, wherein the aforementioned materials also exhibit considerable promise. Pizzi et al. employed MS simulations to examine As- and Sb-based MOSFETs, and their findings indicated that ITRS benchmarks were met, thereby supporting the possibility of utilizing these valuable theoretical insights for further experimental research on these materials.\cite{pizzi_performance_2016}

\begin{figure*}
    \centering   \includegraphics[width=0.9\linewidth]{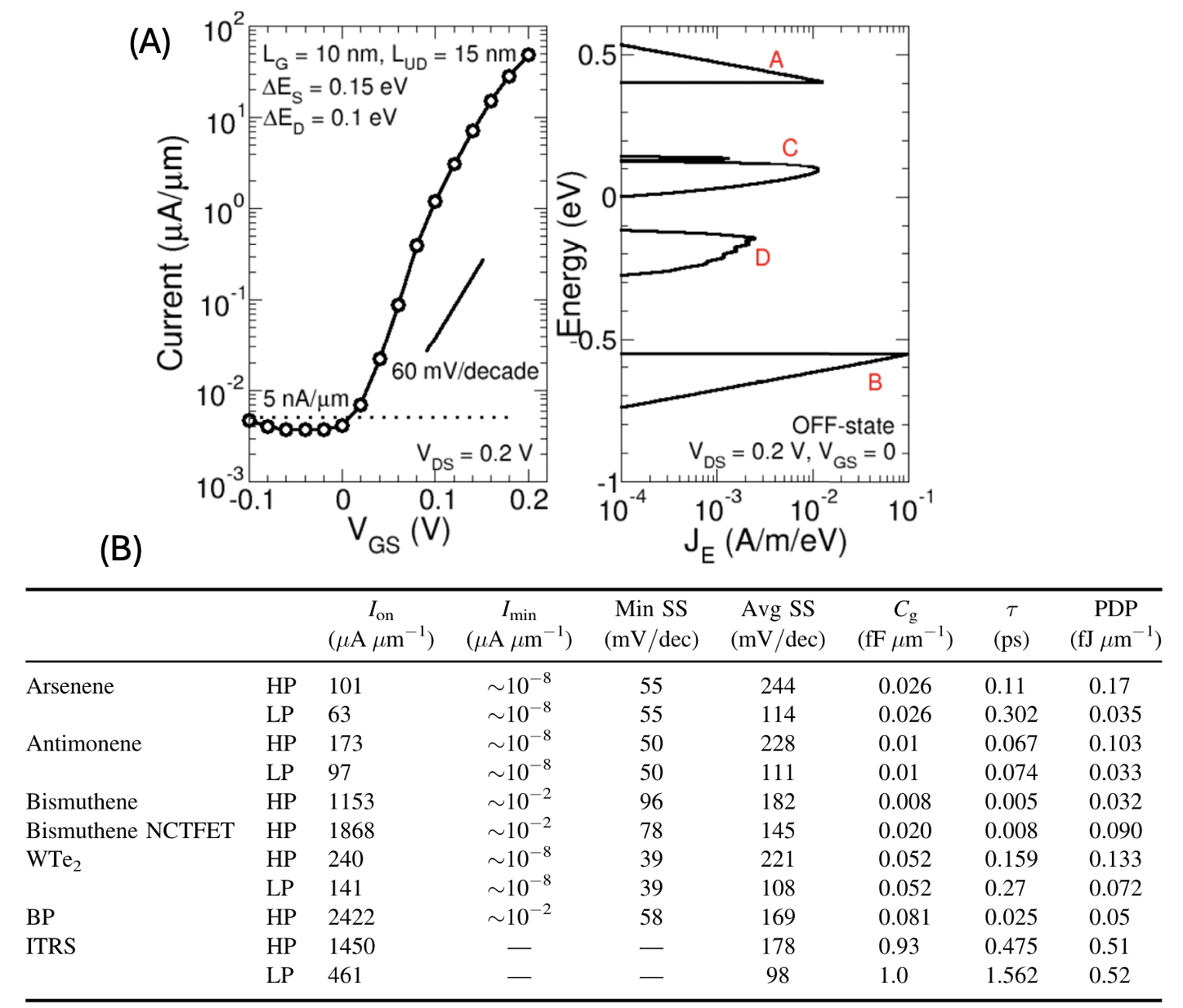}
    \caption{
\textbf{(A)} Left: Transfer characteristic of the n-type 2QL Bi\(_2\)Se\(_3\) TFET at \(V_{DS} = 0.2\,V\) and room temperature, showing an effective subthreshold swing of 50 mV/decade over 4 orders of magnitude. Right: Current spectrum in the OFF-state, demonstrating effective management of all four leakage components. Reproduced with permission from Q. Zhang et al., IEEE Electron Device Lett. 35, 129–131 (2014). Copyright 2014 IEEE.\cite{zhang_two-dimensional_2014}
\textbf{(B)} Benchmark comparison of ballistic device performances of ML group V-ene TFETs and ML bismuthene NCTFETs for high-performance (HP) and low-power (LP) applications, matched against the ITRS 2013 requirements. Also includes performance data for the ML WTe\(_2\) TFET and ML BP TFET. Reproduced with permission from H. Li et al., Semicond. Sci. Technol. 34, 085006 (2019). Copyright 2019 IOP Publishing.\cite{li_negative_2019}}
    \label{fig:Figure6}
\end{figure*}

\section{Machine Learning Methods}
While MS simulations offer comprehensive insights by modeling physical phenomena at scales ranging from atomic to device levels, they are not without significant limitations.\cite{shilko2015overcoming} The integration of quantum mechanical models into MS simulations places a considerable computational burden on the system, which is a necessity for TFETs. This can present a significant challenge to the rapid iteration of devices.\cite{polat_modified_2021} Furthermore, integrating simulations across quantum and classical regimes introduces additional complexity due to the disparate physical models and assumptions inherent to each domain.\cite{peng_multiscale_2021} In instances where high-throughput screening of materials and design parameters is necessary, the computational burden associated with MS simulations represents a significant challenge. Furthermore, the identification of anomalies, such as material defects, is not a straightforward process and frequently necessitates the implementation of extensive simulation customization. The application of ML methods offers a promising avenue for addressing these challenges. They are capable of processing intricate data sets and delivering predictions with greater expediency than MS simulations. In other words, they are capable of modeling high-dimensional spaces in a more efficient manner than MS simulations. With sufficient training, ML algorithms are capable of identifying intricate patterns and relationships in data that are beyond the computational capabilities of MS simulations.\cite{D1CS00503K} Another notable advantage of ML is its scalability,\cite{sharma2022data} which enables the handling and analysis of vast amounts of data with greater efficiency than MS simulations. This is particularly advantageous when investigating novel TFET materials and structures, where the design space can be extensive.

Deep learning (DL) is a high-dimensional method and a subset of ML that employs numerous layers and parameters.\cite{sarker2021deep} In contrast, ML methods such as support vector machines (SVMs), random forests (RF), and gradient boosting machines (GBMs) are lower-dimensional models that serve as data science tools rather than performance prediction methods. They can ingest large datasets and contribute to the design aspects of TFETs, specifically by sifting through various design parameters and effects of device performance to prioritize the parameters that should be of focus. This can assist in streamlining the design process of these devices; however, these methods are not as powerful as DL in predicting device performance. Consequently, ML methods are increasingly becoming essential tools in the TFET domain. They facilitate the acquisition of more rapid, scalable, and frequently more intricate insights into device performance, thereby offering a valuable addition to the insights provided by traditional MS simulations. This section will summarize the essential ML techniques pertinent to TFETs, highlighting their role in enhancing the efficiency and effectiveness of TFET design and performance prediction (Table III).

\begin{table*}[ht]
\centering
\caption{Comparison of common ML methods and their suitability for TFET design and/or prediction.}
\resizebox{\textwidth}{!}{
\begin{tabular}{c|p{4cm}|p{4cm}|p{4cm}|p{4cm}}
\toprule
  & \textbf{Neural Networks} & \textbf{GBMs} & \textbf{RF} & \textbf{SVMs} \\ 
\midrule
\textbf{Pros} & 
\raggedright Models high-dimensional spaces. Complex pattern recognition and prediction.
& 
Iteratively refines predictions. Good for subtle influences in performance.
& 
\raggedright Handles various data types. Good for categorization. Less prone to overfitting.
& 
Generalizes well. Avoids overfitting. Good for small/medium datasets.
\\ 
\midrule
\textbf{Cons} & 
\raggedright Requires substantial training data. Computationally expensive, needs expert tuning.
& 
More prone to overfitting. Requires careful tuning. Computationally intensive.
& 
\raggedright Sensitive to changes in the training set. Less efficient with high-dimensional data.
& 
Binary. Ineffective for multi-class problems. Struggles with large datasets.
\\ 
\bottomrule
\end{tabular}
}
\label{tab:ml-comparison}
\end{table*}

\subsection{Artificial Neural Networks (Deep Learning)}
The application of neural networks, a technique used within DL, represents an optimal approach to the design of TFETs, facilitating the optimization of both TFET architectures and materials. Such techniques are effective in forecasting performance metrics, including on/off ratios and SS.  Artificial neural networks in DL are computational models that emulate the structural organization of the brain's neural networks.\cite{abdolrasol_artificial_2021} These networks are composed of interconnected layers of nodes that process and relay information. They are particularly noteworthy in the field of computational modeling. A typical network comprises three layers: an input layer, multiple hidden layers, and an output layer.\cite{8628144} Each node within the network assigns weights to inputs and utilizes an activation function to generate an output. DL is particularly adept at processing complex data, such as images, due to its ability to learn diverse data features at varying levels of complexity through its multiple layers.\cite{choi_introduction_2020} In the context of TFETs, neural networks demonstrate a particular aptitude for discerning how alterations in design may influence device performance. This is achieved through the analysis of data pertaining to material types, geometries, and other design parameters. Methods such as RF, GBMs, and SVMs are effective at identifying which parameters are most influential in a TFET's design, thereby guiding the prioritization of design strategies. However, they may not fully capture the range of complexities involved in performance prediction with the same efficacy as DL. Conversely, DL networks are capable of optimizing design parameters by identifying patterns within the data and making accurate predictions about device performance. This dual capability is due to their ability to abstract different levels of features from raw data, learn from them, and make predictions based on a deep understanding of the underlying relationships.\cite{guo_deep_2016} Consequently, neural networks serve as a comprehensive tool for TFET development, offering advantages in both optimizing design parameters and predicting device performance with a higher degree of sophistication and accuracy than their lower-dimensional counterparts.

Wang et al. employed DL to address the limitations of Si-TFETs, namely low on-state currents and significant ambipolar leakage.\cite{wang_optimization_2022} They achieved this by proposing a GeSi/Si heterojunction double-gate TFET with a T-channel hetero-gate dielectric structure. The DL model was able to achieve high predictive accuracy and re-emphasizes the opportunity to predict performance from given design parameters in a more efficient and direct optimization process. Furthermore, this model incorporated both forward and inverse design principles, which suggest optimal device structures based on targeted performance goals. This enables custom TFET engineering for devices with specific applications.

In a recent study, Choudhary et al. employed an ML technique, the Atomistic Line Graph Neural Network (ALIGNN), in conjunction with DFT for the design of 2D van der Waals heterostructures.\cite{choudhary2023efficient} A total of 674 non-metallic two-dimensional materials were subjected to analysis, with the objective of creating 226,779 potential heterostructures. The results yielded insights into the most prevalent types of heterostructures, which were identified as type II and type III, the least common. This approach enabled the extraction of insights into chemical trends and potential applications in photocatalysis and high work function metal contacts for devices. Consequently, the deployment of ML tools to predict band alignment information can markedly accelerate the material selection process for device applications. The integration of ML in this context reiterates the accelerated development of device design and optimization, and how it can also enable a more targeted exploration of a vast device design space for 2D materials.

Inspired by physical principles, Li et al. put forth a neural network methodology, for modeling TFET devices.\cite{li_physics-inspired_2016} This method addresses the limitations of traditional multilayer perceptron (MLP) neural networks, which often fail to incorporate the physical principles that govern the device's operation, resulting in models that exhibit unphysical behavior. In contrast, physics-inspired neural network (Pi-NN) assures the precision and efficacy of generated models by integrating the fundamental physics of the TFET device into its neural network architecture. This is accomplished by processing disparate input variables through discrete subnetworks, which are configured to reflect particular physical effects on device performance. To illustrate, the method proposed by Li et al. employs tanh and sigmoid activation functions in various network components to emulate the physical response of TFETs to alterations in drain-source voltage (V\textsubscript{DS}) and gate voltage (V\textsubscript{TG}).\cite{li_physics-inspired_2016} This is demonstrated by the model's ability to ensure that the current is equal to zero when V\textsubscript{DS} is equal to zero, which illustrates a well-behaved I\textsubscript{D}-V\textsubscript{DS} relationship around V\textsubscript{DS} = 0 and excellent subthreshold region fitting (see Fig.7). This approach facilitates the generation of more refined and precise transfer characteristics (I-V curves) from discrete data points, while simultaneously reducing the complexity of the underlying model. The Pi-NN method employs a relatively smaller number of parameters (7 neurons and 20 parameters in total), thereby prioritizing enhanced computational efficiency and performance. It has the potential to facilitate more rapid and reliable design and optimization of TFETs and other electronic devices by integrating the depth of physical modelling with the flexibility of ML approaches.

\begin{figure*}
    \centering \includegraphics[width=0.5\linewidth]{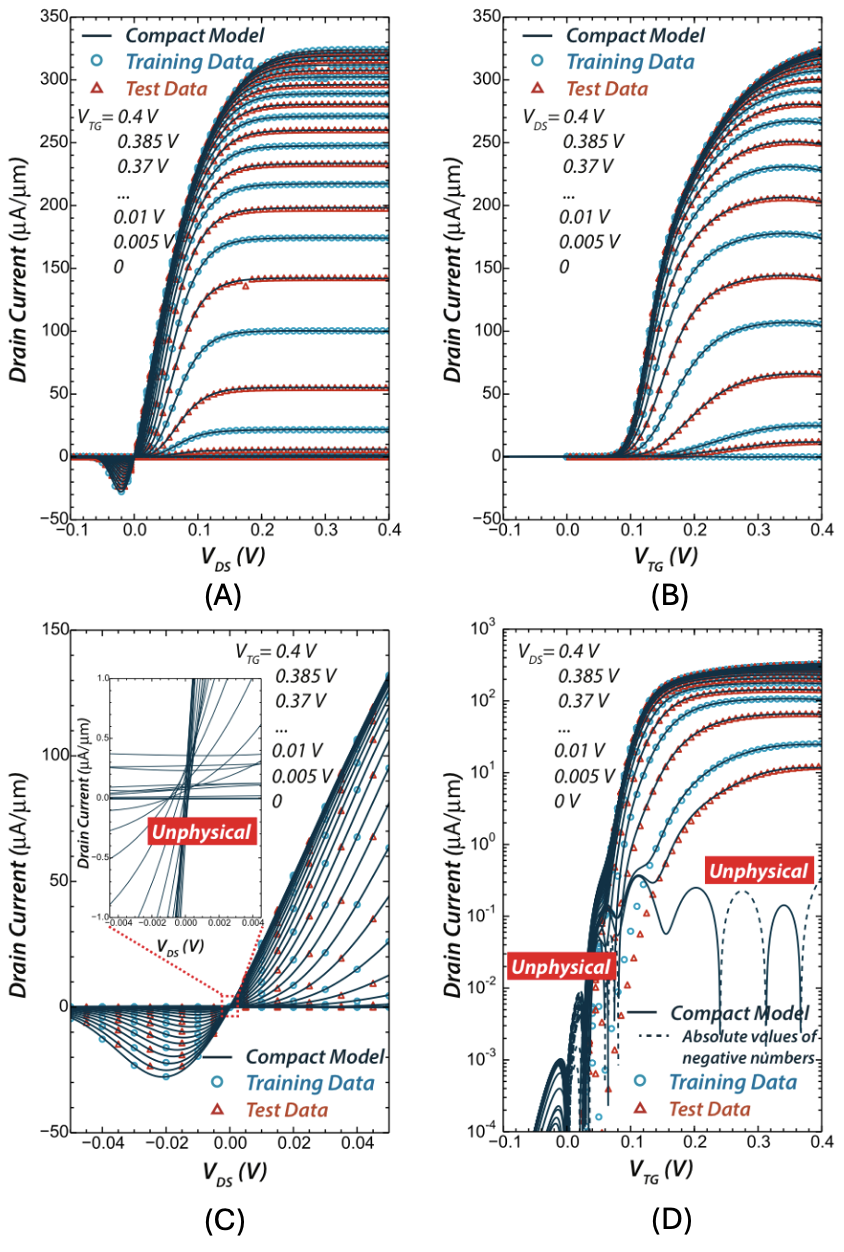}
    \caption{\textbf{(A)} I\textsubscript{D} versus V\textsubscript{DS} at different V\textsubscript{TG} values. \textbf{(B)} I\textsubscript{D} versus V\textsubscript{TG} at different V\textsubscript{DS} values in linear scale. \textbf{(C)} I\textsubscript{D} versus V\textsubscript{DS} at different V\textsubscript{TG} values around V\textsubscript{DS} = 0; the embedded plot shows unphysical I\textsubscript{D}–V\textsubscript{DS} relationships around V\textsubscript{DS} equals 0. \textbf{(D)} I\textsubscript{D}  versus V\textsubscript{TG}  at different V\textsubscript{DS} values in semilog scale; unphysical oscillation of ID around zero appears in the subthreshold region and when V\textsubscript{DS} = 0. Reproduced with permission from M. Li et al., IEEE J. Explor. Solid-State Comput. Devices Circuits 2, 44–49 (2016). Copyright 2016 IEEE.\cite{li_physics-inspired_2016}}
    \label{fig:Figure7}
\end{figure*}

Wu and Guo et al. presented an ML-based framework that employs DL with the objective of streamlining quantum device simulations, with a particular focus on ferroelectric tunnel junctions (FTJs).\cite{wu_speed_2020} The results demonstrated the efficacy of the DL technique in reducing the feature size of device properties while maintaining a sparse representation, thereby retaining key information.  Regression algorithms, specifically Kernel Ridge Regression, show high prediction accuracy with a small training dataset. The framework's computational efficiency is evidenced by a prediction speed that is 10,000 times faster than that of NEGF simulations. The methodology included employing FL for dimensionality reduction, implementing regression algorithms to establish parameter-property mapping, and refining the relationship through feature engineering. It illustrates how applying ML model holds superiority over MS simulations. Nevertheless, in comparison with the Pi-NN model, the integration of fundamental physics into ML models results in a more expeditious and efficacious design and optimization of TFET technology.

\subsection{Other ML Methods}

\subsubsection{Support vector machines}
Among the ML methods discussed, SVMs are distinguished by their relative simplicity.\cite{sheykhmousa_support_2020} The principal objective of this method is to identify a hyperplane within a multidimensional space that can effectively segregate data points into distinct categories.\cite{bhavsar2012review} SVMs are particularly well-suited to small and medium-sized datasets, as they are highly effective at producing models that generalize well and avoid overfitting when tuned correctly. It should be noted, however, that the process of proper tuning can present a significant challenge. SVMs are inherently designed for binary classification, which can render them less practical for multi-class problems that require supplementary techniques. In the context of TFET design, SVMs are particularly effective when the data relationships are evident and the design landscape is more comprehensible. However, their limitations become apparent when confronted with large datasets or tasks that necessitate navigation through complex, high-dimensional data spaces.\cite{bhavsar2012review} Given their relative simplicity in comparison with deep learning and other advanced machine learning approaches, SVMs are recommended for use in the preliminary phases of the TFET design process. They provide a robust foundation for preliminary exploration of the design space, offering a more straightforward computational alternative before transitioning to more sophisticated, computationally intensive models for further refinement.

Murugathas et al. employed SVMs to predict the performance parameters of carbon nanotube (CNT) bundle network FETs under liquid-gated conditions.\cite{murugathas_prediction_2022} A total of 119 devices were examined to explore the role of CNT junctions in electrical conduction and gating. The input parameter was the CNT bundle density, which was measured using atomic force microscopy (AFM) images. The target parameters were the on/off current and threshold voltage of FETs (see Fig.8(A-C)). The on-current was predicted with greater than 90\% accuracy, while the off-current and threshold voltage were predicted with approximately 82\% and 77\% accuracy, respectively. Correlation issues and inaccuracy were observed, which were affected by other parameters, such as network composition. Nevertheless, the effectiveness of SVMs in predicting electronic parameters was demonstrated, despite these issues. It should be noted that SVMs were able to achieve good results in this study due to the presence of strong correlations in the dataset, specifically between on-current and CNT bundle density. However, the accuracy of the model was found to be significantly influenced by even slight variations and complexities in the data.

In another study, Bian et al. employed SVMs and CNT to develop a carbon nanotube-based FETs (NTFETs) decorated with metal nanoparticles for the detection and discrimination of purine compounds.\cite{bian_machine-learning_2019} By applying SVMs and linear discriminant analysis ML methods to the NTFET data, they demonstrated a 93.4\% accuracy rate with a reduced feature set of only 11, which outperformed the 95\% accuracy achieved through a linear discriminant analysis with 48 features. The NTFET characteristics of transconductance, threshold voltages, and minimum conductance were identified as the primary sensing descriptors for classification. Xu et al. presented another example of the use of SVMs for the modeling of SiC based metal–semiconductor FETs, which comprise a cap layer, channel layer, and a buffer layer on a 4H-SiC substrate.\cite{xu_modeling_2007} The input parameters were the gate-source voltage, the drain-source voltage, and the operational frequency. The output parameters were the S-parameters. The SVMs demonstrated satisfactory accuracy in predicting FET performance, as evidenced by MSE values in the range of 1.83e-5 to 2.60e-3 and correlation coefficient (R) values from 0.971 to 0.997 for the training data set and R values from 0.915 to 0.991 for the testing data set.\cite{xu_modeling_2007}

For TFET design, the aforementioned implications serve to reinforce the suggestion that SVMs be employed for the initial stages of TFET design, particularly for tasks such as feature selection and the optimization of design parameters. The model's capacity to operate with a more condensed feature set while maintaining high accuracy makes SVMs an attractive option for streamlining the design process. The model's capacity to achieve this without the necessity for extensive empirical model.

\subsubsection{Random Forest}
RF regression (RFR) is a ML method particularly well-suited to the dimensionality of TFET design parameter optimization, as opposed to performance prediction. RFR functions by constructing a multitude of decision trees during the training phase and subsequently aggregating their predictions. The method's resilience to overfitting and versatility with different types of data make it an ideal tool for identifying which parameters are most crucial in the TFET design process. It has been demonstrated to be applicable to a diverse range of prediction problems with a limited number of parameters to be tuned, and it is suitable for smaller datasets and high-dimensional feature spaces for categorization\cite{biau_random_2016,more2017intelligent, belgiu_random_2016}. However, it has been observed to be sensitive to minor alterations in the training dataset and may be less resilient to overfitting than GBMs.\cite{boateng_basic_2020} This method is highly adept at understanding the manner in which various design parameters influence key performance metrics, including on/off current ratios and threshold voltages. Nevertheless, RFR, in conjunction with SVMs and GBMs, is best utilized for the prioritization of design alternatives. These data science tools, while formidable, do not rival the predictive capacity of deep learning models and are recommended for the guidance of design rather than the prediction and optimization of device performance.

\begin{figure*}
    \centering
    \includegraphics[width=1\linewidth]{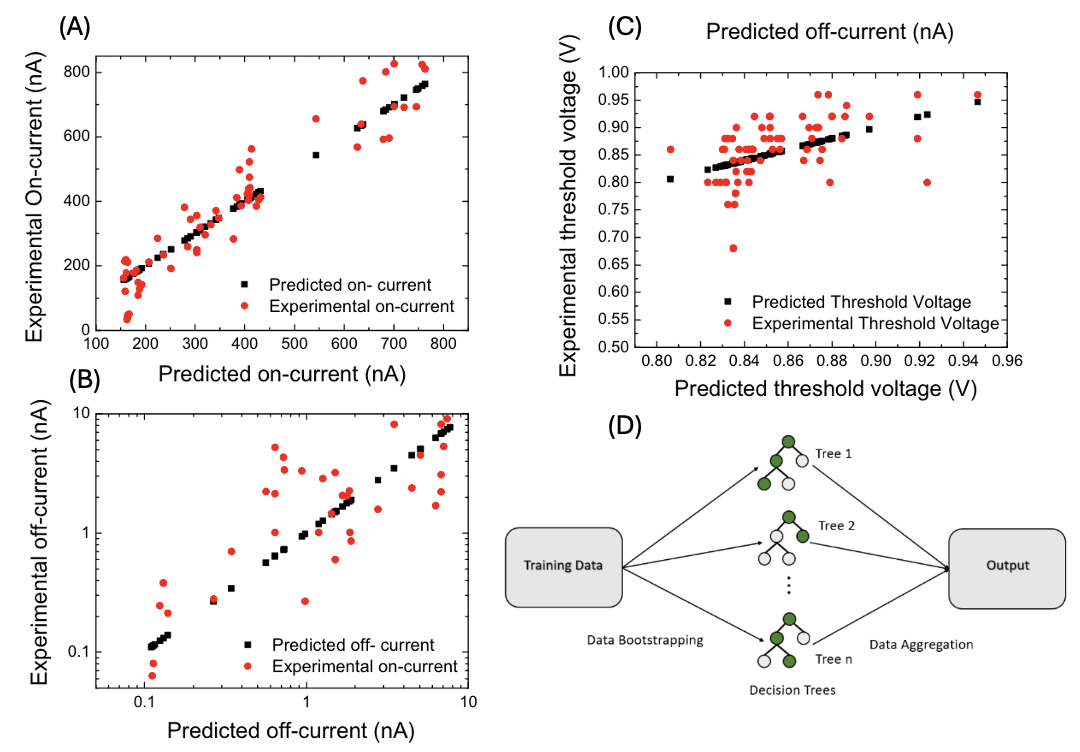}
    \caption{Comparison between the observed and predicted values of \textbf{(A)} on-current, \textbf{(B)} off-current and \textbf{(C)} threshold voltages of a liquid gated CNT network FETs by SVM model. Reproduced with permission from T. Murugathas et al., 2022 IEEE International Conference on Nanoelectronics, Nanophotonics, Nanomaterials, Nanobioscience \& Nanotechnology (5NANO), 1–5 (2022). Copyright 2022 IEEE.\cite{murugathas_prediction_2022} \textbf{(D)} Schematic representation of the working of a random forest algorithm. Reproduced with permission from R. Nirosha et al., 2023 International Conference on Recent Advances in Electrical, Electronics, Ubiquitous Communication, and Computational Intelligence (RAEEUCCI), 1–6 (2023). Copyright 2023 IEEE.\cite{nirosha_analysis_2023}}
    \label{fig:Figure8}
\end{figure*}

In a recent study, Nirosha et al. employed a RF model (Fig.8D), to assess the role of contact resistance (Rc) in organic thin film transistors (OTFTs). Subsequently, the researchers optimized the prediction of Rc behavior.\cite{nirosha2023recent} Although not specifically focused on TFET devices, their study offers valuable insight into the potential utility of the RFR approach. The model is trained using labeled data to predict Rc values based on specified inputs, including dielectric constant, trap concentration, temperature, and channel length. The model is notable for its accuracy and efficiency in both classification and regression tasks. The efficacy of RFR in addressing intricate, non-linear relationships between input variables and Rc in OTFTs exemplifies the potential of ML to elucidate and model the complex interdependencies inherent to TFETs. This allows for a better understanding of how different factors affect the overall performance of the device, which in turn can inform targeted improvements. The model demonstrated high accuracy rates for a range of parameters that affect Rc, thereby illustrating its ability to provide reliable insights into Rc optimization. 

In designing TFETs, this signifies the capacity to anticipate device behavior in response to alterations in material properties, geometries, and environmental conditions. As a result, optimal design configurations for desired operational characteristics can be identified. For example, Akbar et al. utilized RFR to predict the performance of TFETs,\cite{akbar2021vlsi} a methodology that is analogous to that employed by Nirosha et al.\cite{nirosha2023recent} in their analysis of Rc. Their findings illustrate the efficacy of the model in accurately predicting key TFET metrics, including on-current, off-current, and SS. Other significant findings include the model's ability to identify the most influential factors affecting performance by analyzing various device parameters, which can assist in optimizing the design of TFETs. This approach can significantly enhance the design process by streamlining the development and guiding decisions at early stages. It does so by having the ability to analyze large datasets and identify critical design parameters, as well as providing rapid predictions on device performance, which allows for quicker iterations and optimizations. Furthermore, the employment of ML techniques also contributes to a significant reduction in computational costs.  

\subsubsection{Gradient Boosting Machines (GBMs) and XGBoost}
Extreme gradient boosting (XGBoost) is a specific type of GBMs. It is a supervised classification ML algorithm that has been trained using a Pearson correlation coefficient and an important feature metric in order to evaluate the performance of the training features. Both GBMs and XGB represent a refinement of RFR. They are designed to enhance model accuracy by iteratively addressing prediction errors. These methods employ ensembles of simple decision trees, which enable incremental refinement of predictions.\cite{natekin_gradient_2013} Although they operate in a manner analogous to RFR, GBMs and XGBoost are distinguished by their capacity to train more effectively, which is particularly advantageous when optimizing TFET design parameters. The GBM method is particularly effective in addressing the most challenging data points, thereby facilitating the production of increasingly accurate models. However, if not properly calibrated, GBMs have the potential to overfit, thereby requiring greater computational resources. Additionally, they are susceptible to minor alterations in the training dataset, prompting the generation of a new tree.\cite{bentejac_comparative_2021} Conversely, RFR provides a robust basis for assessing the significance of parameters. It is user-friendly, requires minimal tuning, and maintains robustness against overfitting while reliably yielding good performance. However, GBMs and XGBoost go further by necessitating fine-tuning. While they may be more susceptible to overfitting if neglected, they compensate by delivering more precise results swiftly. Thus, GBMs and XGBoost are particularly adept at quickly pinpointing crucial design elements that can enhance current efficiency and switching behaviors in TFETs, underscoring their potential to accelerate the design and optimization process in TFET technology.

Chen et al. employed the XGB method to examine a multitude of potential heterojunction candidates with the objective of identifying a high-performance 2D vdW metal-semiconductor heterojunction.\cite{chen_accelerated_2022} The ML screening process identified six candidates (BTe–NbSe\textsubscript{2}, Al\textsubscript{2}SO–Zn\textsubscript{3}C\textsubscript{2}, iAl\textsubscript{2}SO–Zn\textsubscript{3}C\textsubscript{2}, GaSe–NbS\textsubscript{2}, GaSe–NbSe\textsubscript{2}, and GeSe–VS\textsubscript{2}) from 1092 candidates that exhibited Ohmic contacts and high tunneling probabilities, which are essential for optimizing contact resistance and enhancing device performance (see Table \ref{tab:chen_table}). More importantly, the XGB method is executed in less than 5 seconds and is demonstrably more efficient than traditional first-principles calculations in terms of both time and cost. This evidence supports the assertion that machine learning is an effective approach for screening materials and that unsupervised assisted algorithms can address the challenge of data scarcity in predicting the behaviors of complex dynamical systems.

\begin{table}[ht]
    \centering
    \caption{Minimum Interfacial Distance ($d_{min}$), Binding Energy ($E_b$), Work Function of 2D Metals ($W_m$), Schottky Barrier Height ($\Phi_{SB}$), and Tunneling Probability (TP) of 27 vdW Heterostructures That Achieved Ohmic Contact. Reproduced with permission from A. Chen et al., Chem. Mater. 34, 5571–5583 (2022). Copyright 2022 American Chemical Society.\cite{chen_accelerated_2022}.}
    \begin{tabular}{|l|c|c|c|c|c|}
        \hline
        \textbf{Systems} & \textbf{$d_{min}$/($\text{\AA}$)} & \textbf{$E_b$/(eV)} & \textbf{$W_m$/(eV)} & \textbf{$\Phi_{SB}$/(eV)} & \textbf{TP/(\%)} \\
        \hline
        BTe–NbS$_2$ & 3.88 & –0.0179 & 6.12 & –0.0615 & 2.5960 \\
        BTe–VS$_2$ & 3.64 & 0.0857 & 5.98 & –0.2446 & 3.7205 \\
        AlO–VSe$_2$ & 3.95 & 0.0810 & 5.39 & –0.0563 & 3.1002 \\
        AlSe–g & 3.96 & –0.0360 & 4.51 & –0.3683 & 3.6162 \\
        Al$_2$SeO–g & 3.95 & –0.0266 & 4.51 & –0.0200 & 3.3140 \\
        Al$_2$SeO–NbS$_2$ & 3.99 & –0.0569 & 6.12 & –0.0031 & 1.1781 \\
        Al$_2$SeO–VS$_2$ & 3.91 & –0.0382 & 5.98 & –0.0403 & 1.1768 \\
        GaO–NbS$_2$ & 3.90 & 0.0374 & 6.12 & –0.9749 & 0.7864 \\
        GaO–NbSe$_2$ & 3.71 & 0.0925 & 5.42 & –1.1058 & 1.6569 \\
        GaO–TaS$_2$ & 3.91 & 0.0350 & 5.95 & –0.8536 & 0.6503 \\
        GaO–TaSe$_2$ & 3.98 & 0.1062 & 5.41 & –1.1026 & 0.9843 \\
        GaO–VS$_2$ & 3.99 & –0.0201 & 5.98 & –0.9002 & 0.4457 \\
        GaO–VSe$_2$ & 3.96 & 0.0073 & 5.39 & –0.0878 & 0.6103 \\
        GaSe–g & 4.27 & 0.0640 & 4.51 & –0.1232 & 2.8049 \\
        GaSe–NbS$_2$ & 3.15 & 0.0214 & 6.12 & –0.4552 & 16.5614 \\
        GaSe–NbSe$_2$ & 3.16 & 0.2820 & 5.42 & –0.3176 & 31.3775 \\
        GaTe–g & 3.85 & –0.0531 & 4.51 & –0.2456 & 8.4528 \\
        Ga$_2$SeO–g & 3.61 & –0.0168 & 4.51 & –0.1289 & 4.0458 \\
        Ga$_2$SeO–NbS$_2$ & 3.71 & –0.0370 & 6.12 & –0.0565 & 1.4988 \\
        Ga$_2$SeO–TaS$_2$ & 3.96 & –0.0336 & 5.95 & –0.0267 & 0.9770 \\
        Ga$_2$SeO–VS$_2$ & 3.75 & 0.0214 & 5.98 & –0.2036 & 1.1305 \\
        Ga$_2$SeO–VSe$_2$ & 3.72 & –0.0385 & 5.39 & –0.0191 & 2.5033 \\
        Ga$_2$SSe–g & 3.88 & –0.0386 & 4.51 & –0.4859 & 3.0206 \\
        InS–g & 3.45 & –0.0189 & 4.51 & –0.1327 & 3.4158 \\
        InS–NbS$_2$ & 3.06 & 0.1050 & 6.12 & –0.3443 & 11.4793 \\
        GeS–g & 4.07 & 0.0174 & 4.51 & –0.1582 & 10.3199 \\
        GeSe–VS$_2$ & 3.38 & 0.0580 & 5.98 & –0.0050 & 53.5366 \\
        \hline
    \end{tabular}
    \label{tab:chen_table}
\end{table}

Although there has been limited investigation into the use of GBMs and XBG ML methods for TFET device simulation, other applications of this method indicate that it is a highly useful model for specific applications. In other words, the advantage of employing this methodology for automated design space exploration and parameter optimization, as well as for efficient and accurate performance prediction, is evident. For example, the study by Wang and Ross demonstrated the use of the ML model in predicting travel mode choices.\cite{wang2020travel} The objective of this study was to determine the relative performance of the XGBoost algorithm in comparison to other models, such as RFR, in the context of transportation data. The XGB model was found to have superior accuracy in predicting travel mode choices, thereby demonstrating its strength in dealing with complex, non-linear relationships with data. The ability to do so is of significant benefit for the exploration of new materials and architectures, as evidenced by the literature.  In addition to its accuracy, the model was also able to provide insights into the relative importance of the variables that influence travel mode decisions. This is beneficial when applied to TFET design, as it allows for the prioritization of optimization efforts to enhance device efficiency and effectiveness.

\subsubsection{Future Directions}
Given the advancements in TFET technology and the need for precision in device modeling, Pi-NN methods are recommended for further development. Their high-dimensional processing power is already a significant asset, yet their true potential lies in integrating fundamental quantum mechanical principles directly into the neural network framework. This integration is key, capturing the nuances of TFET operation that traditional neural networks may overlook, making Pi-NNs particularly valuable. For exploring the design space, RF is advantageous when dealing with large, complex datasets and when seeking to understand broad trends. RF requires less computational power compared to more complex models and offers easier fine-tuning, although it is less suitable for high-dimensional data. It serves as a solid starting point for initial explorations when the relationships between design parameters and performance are not yet fully comprehended. When the dataset is smaller and the design parameter relationships are intricate, necessitating detailed fine-tuning, GBMs are suggested. GBMs are more resource-intensive but can deliver enhanced performance in such scenarios. They’re ideal for optimizing design parameters with subtle impacts on TFET device performance. SVMs are ideal for small to medium-sized datasets. They are simpler models that might struggle with large volumes of data but can be very effective in well-mapped design spaces with stable, clear-cut relationships. That is, they can define clear boundaries in design optimization challenges and offer clarity when deciding on the best path forward for TFET designs.

\section{Conclusion}
Through our thorough discussion of the application of various 2D materials for TFET design, it is evident that each material group offers specific advantages depending on the application. While direct bandgap materials are preferred for TFET devices, they may not be suitable for traditional transistors, making it crucial to consider the intended application when selecting materials. BP, with its anisotropic properties, tuneable bandgap, and potential for a low SS, is ideal for low-power applications. Group III-V materials are generally well-suited for TFETs, offering high on-current and efficient tunneling, making them ideal for high-speed, low-voltage applications. TMDs, with their excellent electrostatic control and direct bandgap in monolayer form, are best for ultra-thin body TFETs and high-speed switching applications. Additionally, MS simulations can identify optimal TFET designs without the cost and time associated with experiments. However, integrating quantum models into MS simulations is computationally intensive. In contrast, ML methods efficiently model high-dimensional spaces, making them particularly effective for exploring novel TFET materials and structures. Within our discussion of the ML methods, we’ve also summarized the suitability of each ML method for simulating TFET performance. This combination of material selection and advanced simulation techniques is essential for optimizing TFET performance across various applications.

\begin{acknowledgments}
This work was supported by the National Science Foundation Future Manufacturing Research Grant Program (NSF CMMI-2037026).
\end{acknowledgments}

\section*{Author Declarations}
\subsection*{Conflict of Interest}
The authors have no conflicts to disclose.

\section*{Author Contributions}
\textbf{Chloe Isabella Tsang}: Conceptualization (lead); Investigation (lead); Writing original draft (lead); Writing– review \& editing (lead); Visualization (lead). \textbf{Haihui Pu}: Investigation (lead); Writing–original draft(supporting). Writing– review \& editing (lead); Visualization (lead). \textbf{Junhong Chen}: Conceptualization (supporting); Investigation (supporting); Writing– review \& editing (supporting); Visualization (supporting).

\section*{Data Availability Statement}
Data sharing is not applicable as no new data were created or analyzed in this review paper.

\bibliography{MAIN, RSC_, pericles_1521409534}
\end{document}